\documentclass[12pt]{iopart}
\usepackage{graphicx}
\usepackage{caption}
\usepackage{cite}
\usepackage{subcaption}
\usepackage{amsmath}
\usepackage{mathtools}
\usepackage{algorithm}
\usepackage{algpseudocode}
\usepackage{bbold}
\usepackage{cleveref}
\usepackage{xcolor}

\usepackage[margin=0.95in]{geometry}

\begin{document}

\title[Near-Term Distributed Quantum Computation]{Near-Term Distributed Quantum Computation using Mean-Field Corrections and Auxiliary Qubits}

% \author{Abigail McClain Gomez}
% \address{Department of Physics, Harvard University, Cambridge, Massachusetts 02138, USA}
% \vspace{8pt}
% \address{NVIDIA, Santa Clara, CA 95051, USA}
% \ead{amcclain@g.harvard.edu}

% \author{Taylor L. Patti}
% \address{NVIDIA, Santa Clara, CA 95051, USA}
% \ead{tpatti@nvidia.com}

% \author{Anima Anandkumar}
% \address{Department of Computing + Mathematical Sciences (CMS), California Institute of Technology (Caltech), Pasadena, CA 91125, USA}
% \vspace{8pt}
% \address{NVIDIA, Santa Clara, CA 95051, USA}

% \author{Susanne F. Yelin}
% \address{Department of Physics, Harvard University, Cambridge, Massachusetts 02138, USA}

\author[A. McClain Gomez et al.]
       {Abigail McClain Gomez$^{1,2,*}$, Taylor L. Patti$^{2,\dag}$, 
       \\ 
       Anima Anandkumar$^{2,3}$, and Susanne F. Yelin$^{1}$
       \\
       }
\address{$^1$Department of Physics, Harvard University,            Cambridge, MA 02138, USA\\
       $^2$NVIDIA, Santa Clara, CA 95051, USA\\
       $^3$Department of Computing + Mathematical Sciences (CMS), California Institute of 
       \\ \hspace{1pt} Technology (Caltech), Pasadena, CA 91125, USA\\}
\ead{$^*$amcclain@g.harvard.edu, $^\dag$tpatti@nvidia.com}

\begin{abstract}
Distributed quantum computation is often proposed to increase the scalability of quantum hardware, as it reduces cooperative noise and requisite connectivity by sharing quantum information between distant quantum devices. However, such exchange of quantum information itself poses unique engineering challenges, requiring high gate fidelity and costly non-local operations. To mitigate this, we propose near-term distributed quantum computing, focusing on approximate approaches that involve limited information transfer and conservative entanglement production. We first devise an approximate distributed computing scheme for the time evolution of quantum systems split across any combination of classical and quantum devices. Our procedure harnesses mean-field corrections and auxiliary qubits to link two or more devices classically, optimally encoding the auxiliary qubits to both minimize short-time evolution error and extend the approximate scheme's performance to longer evolution times. We then expand the scheme to include limited quantum information transfer through selective qubit shuffling or teleportation, broadening our method's applicability and boosting its performance. Finally, we build upon these concepts to produce an approximate circuit-cutting technique for the fragmented pre-training of variational quantum algorithms. To characterize our technique, we introduce a non-linear perturbation theory that discerns the critical role of our mean-field corrections in optimization and may be suitable for analyzing other non-linear quantum techniques. This fragmented pre-training is remarkably successful, reducing algorithmic error by orders of magnitude while requiring fewer iterations.

\end{abstract}

\noindent{\it Keywords}: Distributed Quantum Computing, Near-term Quantum Computing, Quantum Simulation, Variational Quantum Algorithms
%\maketitle

\section{Introduction} \label{sec:intro}
One prospective trajectory for quantum information hardware is distributed quantum computing \cite{divincenzo_quantum_1999, denchev_distributed_2008, gyongyosi_survey_2019}, the quantum analog of the celebrated classical field \cite{birman_process_1993, attiya_distributed_2004, kshemkalyani_distributed_2008, hajibaba_review_2014}. 
Distributed quantum computing seeks to eliminate the need for large, monolithic quantum computers, which suffer from cooperative noise \cite{cheng_noisy_2023, cuomo_towards_2020}. Instead, large-scale problems will be split among many smaller quantum computers that are in communication with each other via a \textit{quantum interconnect}, a standardized form of quantum communication between remote quantum computing platforms \cite{caleffi_quantum_2018, cacciapuoti_quantum_2020}. 

While the benefits of distributed quantum computing are abundant, many obstacles complicate its realization. For instance, due to the no-cloning theorem \cite{wootters_single_1982}, extensive quantum entanglement would be a required component of quantum interconnects in order to enable non-local operations such as quantum teleportation \cite{divincenzo_quantum_1999, denchev_distributed_2008, ma_quantum_2012, cuomo_towards_2020}. Moreover, fault-tolerant quantum computing would be needed to compute and transmit quantum information between distributed simulators reliably \cite{van_meter_path_2016, cacciapuoti_quantum_2020}. Finally, long coherence times or relatively local topology would be necessary to manage the time delays associated with communication between remote locations \cite{cuomo_towards_2020, qiao_quantum_2022}.  

Nevertheless, the promise of scalability continues to inspire research in various facets of distributed quantum computing. 
Researchers have characterized the compilation of quantum circuits into cohesive network instructions \cite{ferrari_compiler_2021} and devised a language to communicate such instructions more efficiently than conventional circuit diagrams \cite{ying_algebraic_2009}.
Likewise, much work has been done to develop the non-local operations integral to distributed quantum computing, which have been supported with experimental realizations \cite{yimsiriwattana_generalized_2004, lim_repeat-until-success_2005, ma_quantum_2012, liu_distributed_2023}. Other studies have developed algorithms tailored to quantum distributed architectures, including Shor's algorithm, quantum sensing, and combinatorial optimization \cite{yimsiriwattana_distributed_2004, zhang_distributed_2021, saleem_divide_2022, parekh_quantum_2021}, while additional research has focused on the quantum advantage provided by quantum distributed computing \cite{fitzi_quantum_2001, gavoille_what_2009, avron_quantum_2021, censor-hillel_quantum_2022}. Still other research has addressed how to approach distributed algorithm design \cite{beals_efficient_2013, parekh_quantum_2021}, the effect of noise in distributed quantum computing \cite{cirac_distributed_1999}, architecture selection and scalability \cite{van_meter_distributed_2010, van_meter_local_2016, van_meter_path_2016, gyongyosi_scalable_2021}, and resource allocation \cite{sundaram_efficient_2021, cicconetti_resource_2022, ngoenriang_optimal_2022}, particularly to optimize teleportation cost \cite{daei_optimized_2020, houshmand_evolutionary_2020, cuomo_optimized_2023}.

\begin{figure*}[ht]
    \centering
    \begin{minipage}[b]{2.7in}
    \begin{minipage}[h]{.2in}
    \caption*{(a)} \label{subfig:dist-sim}
    \end{minipage}
    \begin{subfigure}[b]{2.5in}
        \centering\includegraphics[width=2.5in]{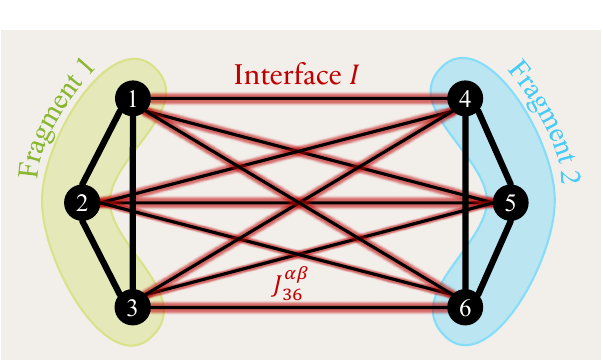}
    \end{subfigure}

    \begin{minipage}[b]{.2in}
    \caption*{(b)} \label{classical-link}
    \end{minipage}
    \begin{subfigure}[b]{2.5in}
        \centering\includegraphics[width=2.5in]{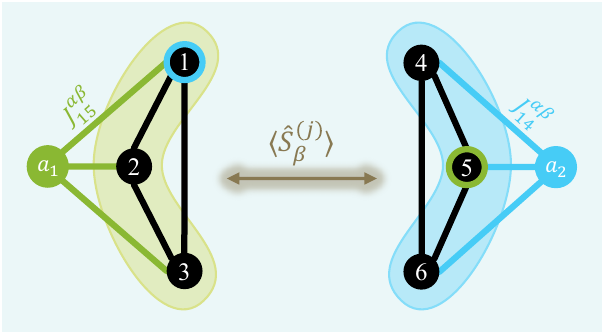}
    \end{subfigure}
    \end{minipage}
    \qquad
    \begin{minipage}[b]{2.7in}
    \begin{minipage}[b]{.2in}
    \caption*{(c)} \label{subfig:quantum-link}
    \end{minipage}
    \begin{subfigure}[b]{2.5in}
        \centering
        \includegraphics[width=2.5in]{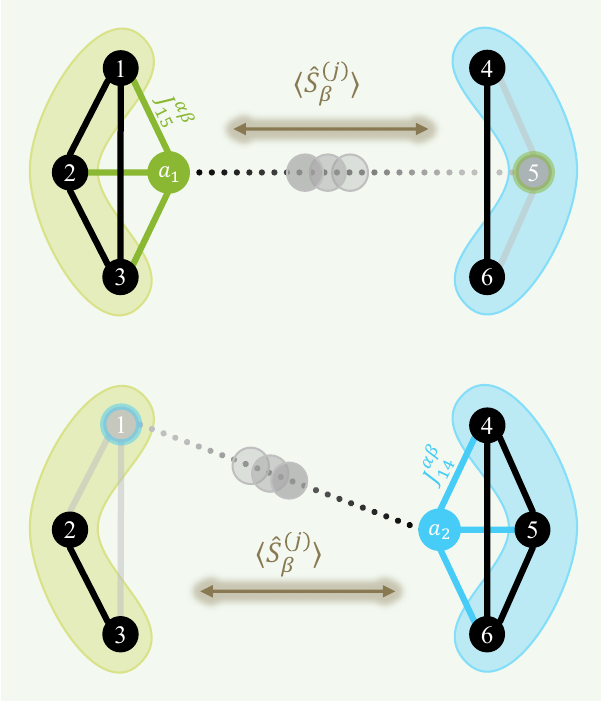}
    \end{subfigure}
    \end{minipage}
    \caption{1a: Diagram illustrating how a 6-qubit system can be split into two fragments. Interactions $J_{ij}^{\alpha \beta}$ are represented by lines between qubits; one label is included for clarity. Interactions that span the two fragments form the interface $I$. 1b: The case where the two fragments are linked via a classical channel. Mean-field measurements $\langle \hat{S}_{\beta}^{(j)}\rangle$ are exchanged classically. One auxiliary qubit $a_{1, 2}$ is included in each fragment's simulation, interacting with the fragment qubits according to a target qubit in the opposite fragment (identified in the figure by a blue / green circle). 1c: The case where the two fragments are linked via a quantum channel. While mean-field measurements $\langle \hat{S}_{\beta}^{(j)}\rangle$ are still exchanged classically, the auxiliary qubits are physically shared between fragments using some form of quantum communication.}
    \label{fig:link-comp}
\end{figure*}

Although the interest in its theoretical application continues to grow, a wide gap remains between much distributed quantum computing research and its physical implementation. Research along a different vein has instead concentrated on applications that are realizable using near-term hardware, stretching the limit of noisy quantum simulators' utility. Although not distributed in the sense of the works discussed above (which assume that the distributed hardware forms a quantum network), these approaches involve small groups of qubits simulated in parallel or in sequence to address a larger problem. Entanglement forging is one such approach \cite{eddins_doubling_2022}, which relies on shifting computation to classical post-processing in order to assemble information from two smaller circuits, thereby halving the maximum circuit size required for the calculation. The quantum tensor network approach uses the framework of tensor networks to identify weakly entangled subgroups and parallelize quantum simulation \cite{barratt_parallel_2021}. Similarly, Quantum Multi-Programming (QMP) takes advantage of the increasing size of available quantum simulators to execute multiple shallow quantum circuits concurrently \cite{das_case_2019, park_quantum_2022}.

In order to bridge the distributed quantum computing paradigm with the capabilities of near and moderate-term hardware, in this manuscript, we design two procedures that approximately link distributed simulators while remaining amenable to small-scale, noisy devices. Our schemes of fragmented quantum simulation explore what problems can be addressed without full information transfer between hardware. First, focusing on the task of time evolution, we partition a system of qubits into subgroups (referred to as \textit{fragments}) that are treated separately. We harness mean-field measurements to inform mean-field corrections \cite{strecka_brief_2015} that link the distinct fragments. These simulations could be executed in parallel on a single simulator (as in QMP \cite{das_case_2019, park_quantum_2022}), outsourced to different simulators (as in distributed computing \cite{divincenzo_quantum_1999, denchev_distributed_2008, ma_quantum_2012, cuomo_towards_2020}), or even simulated using a mixture of classical and quantum resources (as in \textit{heterogeneous} computing \cite{mccaskey_hybrid_2018, britt_high-performance_2017}). We further make use of a limited number of auxiliary qubits to mimic the presence of the qubits located on distant simulators. 

In our first approach to distributed time evolution, we rely on classical communication to transmit partial state information between distant simulators through measurements, omitting a quantum link between devices.
Transmitting incomplete information reduces the generally exponential number of measurements required to relay complete information of a quantum state via a classical channel. For locally interacting systems, the classical fragmentation scheme closely approximates quantities local to each fragment -- including the fidelity of the fragment -- for timescales up to several $1/J$, where $J$ weights the system's interactions. We present a second scheme that is supplemented by limited quantum information transfer, consequently composing an interface of classical and partial quantum information transfer that approximately connects quantum simulators. We show numerically that the limited use of quantum communication significantly extends the scheme's performance to longer evolution times, even for long-range interacting systems. As non-local operations become more available, this technique could be employed in moderate-term distributed applications before a fully connected quantum network is achievable.

Using the same fragmentation framework, we devise a fragmented pre-training approach for variational quantum algorithms, focusing on the variational quantum eigensolver algorithm (VQE) \cite{fedorov_vqe_2022}. The pre-training can be performed classically or using resource-limited hardware, as only portions of the full circuit are considered. For classical MaxCut problem graphs, the pre-training method reduces energy error by various orders of magnitude on average, and requires over an order of magnitude fewer circuit preparations. For transverse field Ising-like models \cite{pfeuty_one-dimensional_1970, stinchcombe_ising_1973} outside of the classical domain, our pre-training scheme maintains a significant advantage in the regime of a small transverse field $h$.

The remainder of the paper is organized as follows. In Section \ref{sec:frag-q-sim}, we first present a fragmented approach to quantum simulation that only involves the classical transfer of partial state information. We further consider an alternate scheme for the case of linking quantum simulators with reduced quantum information transfer through selective qubit shuttling \cite{noiri_shuttling-based_2022} or teleportation \cite{bennett_teleporting_1993, bouwmeester_experimental_1997}, in addition to classical information transfer. In Section \ref{sec:frag-TE}, the performance of each scheme is evaluated for the time evolution of quantum Ising-like spin Hamiltonians \cite{parkinson_introduction_2010}, which are amenable to quantum simulation using trapped ions and Rydberg platforms \cite{kim_quantum_2011, schauss_quantum_2018}. Finally, in Section \ref{sec:frag-circ} we expand the scheme to apply to the optimization of quantum circuits. The use of our fragmentation scheme to assist VQE is evaluated in Section \ref{sec:frag-VQE} \cite{fedorov_vqe_2022}. The role of mean-field corrections in the optimization through the lens of perturbation theory \cite{sakurai_modern_2020} is explored in Sections \ref{sec:solve-maxcut} and \ref{sec:mf-pt}. In Section \ref{sec:mf-pt}, we introduce a non-linear perturbation theory to study mean-field corrected Hamiltonians, analytically formalizing the success of our pre-training approach.

\section{Fragmented Quantum Simulation} \label{sec:frag-q-sim}
In our method of fragmented quantum simulation, we divide a system of $N$ qubits into two or more sub-systems, here referred to as \textit{fragments} (see Fig. \ref{fig:link-comp}a). Each fragment contains some number of qubits $N_{f} < N$, such that $\sum_{f}{N_f} = N$. The fragments are treated separately, but it is possible to approximate the presence of a fragment's \textit{environment}, that is, the qubits outside of a given fragment, through corrective fields and interactions \cite{wouters_practical_2016}. We devise mean-field corrections (described in detail in Section \ref{sec:MF}) \cite{strecka_brief_2015}, which are informed by measurements of a fragment's environment, to actively adjust the state of a fragment. Corrective interactions are mediated by the inclusion of auxiliary qubits within each fragment's simulation, such that $\sum_{f} N_{f+a} > N$, where $N_{f+a} = N_f + N_a$ and $N_a$ represents the number of auxiliary qubits included in fragment $f$. Each auxiliary qubit mimics the behavior of one environment qubit, which we refer to as the \textit{target} qubit for that auxiliary. Each auxiliary qubit interacts with the fragment's qubits according to the same interaction terms as the corresponding target qubit, as prescribed by the original Hamiltonian, enabling entanglement to grow beyond the $N_f$ fragment qubits. 

Fig. \ref{fig:link-comp}b provides an overview of our classically-linked fragmentation scheme, and a detailed diagram is provided in Fig. \ref{fig:fragment-cartoon}. We define a fragment's \textit{interface} $I$ to be the collection of interactions existing in the original Hamiltonian that act between fragment qubits and environment qubits. The combination of auxiliary qubits and mean-field corrections collectively mimics the action of the interface on the fragment. The growth and faithfulness of the entanglement within a fragment will be limited by the number of auxiliary qubits included -- an unavoidable limitation of the scheme -- but the effects of this limitation can be mitigated through judicious fragmentation of the system. Firstly, to mitigate fragmentation error (that is, the error produced by the omission of some system interactions and the resultant reduction of Hilbert space), one can choose to divide the system qubits such that the qubits interacting most influentially with each other are confined to a single fragment. Secondly, it is possible to make an informed choice of target qubit for each auxiliary. This is explored further in Section \ref{sec:aux-selection}.

\subsection{Mean-Field Corrections} \label{sec:MF}
Consider the class of spin models: 
\begin{equation}
    H = -\sum_{\langle i,j \rangle} \sum_{\alpha, \beta} J_{ij}^{\alpha, \beta} \hat{S}^{(i)}_{\alpha} \hat{S}^{(j)}_{\beta} - \sum_{i = 1}^{N} h_{i} \hat{S}^{(i)}_x .
\end{equation}
Here, $\hat{S}^{(i)}_{\alpha}$ and $\hat{S}^{(j)}_{\beta}$ are spin-1/2 spin operators acting on sites $i$ and $j$, where $\alpha, \beta \in \{x, y, z\}$. The coefficient $J_{ij}^{\alpha, \beta}$ gives the strength and sign of the interaction. For concreteness and without loss of generality, we have selected transverse fields $h_{i}$ to point along the x-axis. The Hamiltonian acting strictly within some sub-system $f$ will neglect any operators acting outside of $f$, yielding
\begin{equation} 
    H^{(f)} = -\sum_{\langle i,j \rangle \in f}
    \sum_{\alpha, \beta} J_{ij}^{\alpha, \beta} \hat{S}^{(i)}_{\alpha} \hat{S}^{(j)}_{\beta} -  \sum_{i\in f} h_{i} \hat{S}^{(i)}_x ,
\end{equation}
\noindent the bare Hamiltonian that acts within a fragment $f$ when no corrections are included.

Clearly, the simple exclusion of interactions that span the interface between $f$ and its environment (i.e., the fragmented evolution of $f$ under $H^{(f)}$) will, in general, poorly approximate the evolution of the sub-system under the full Hamiltonian. The fragment qubits will behave as a closed system without external interactions. Although generally these interactions cannot be exactly simulated without modeling all of the system's spins on a single fragment, we introduce a mean-field to partially capture the action of each missing interaction. Mean-field methods have frequently been used to simplify the simulation and study of quantum systems, and statistical physics \cite{hes_mean-field_1996, bobbio_analysis_2008, strecka_brief_2015}. Here, the strength and sign of the introduced mean-field correction is informed by the measurement of the corresponding environment spin, while the correction's axis is determined by that of the corresponding interaction's spin operator that would act within fragment $f$. The resulting mean-field corrected Hamiltonian is given by:
\begin{equation}
    \label{eq:Ham-MF}
    H^{(f)}_{MF} = -\sum_{\substack{\langle i,j \rangle \in   I,\\i \in f}} \sum_{\alpha, \beta} J_{ij}^{\alpha, \beta} \hat{S}^{(i)}_{\alpha} \langle \hat{S}^{(j)}_{\beta} \rangle - \sum_{\langle i,j \rangle \in f} \sum_{\alpha, \beta} J_{ij}^{\alpha, \beta} \hat{S}^{(i)}_{\alpha} \hat{S}^{(j)}_{\beta} - \sum_{i \in f}h_{i} \hat{S}^{(i)}_x .
\end{equation}
The strength and direction of the mean-fields appearing in $H^{(f)}_{MF}$ should be updated regularly to reflect the current state of the environment spins. Physically, this requires regular mean-field measurements of the fragments. Evolution must therefore be reset to the initial state in order to proceed by one time step $dt$, with each new mean-field measurement being stored to progress the evolution. The process of incrementing the time evolution by one time step per simulation is commonly implemented in order to track the time dynamics of an observable \cite{sargent_laser_1978}, resulting in a complexity that scales as $\mathcal{O}(N_t^2)$ in the number of time steps $N_t$.

\subsection{Auxiliary Target Spin Selection} \label{sec:aux-selection}
For nearest-neighbor spin models (e.g., the transverse field Ising model \cite{stinchcombe_ising_1973}), the selection of a target spin for each auxiliary is somewhat trivial, as at most two qubits interact with a fragmented section of the chain. The choice of auxiliary qubit encoding may be unclear for more general systems. Here, we present a method for auxiliary target qubit selection that yields, on average, the optimal auxiliary qubit encoding. Specifically, we consider how auxiliary target selection affects the simulation error to the first non-vanishing order in $dt$. This simulation error arises from the omission of interactions forming the interface of some particular fragment $f$ and the remaining environment spins $E$. The full derivation of the leading error is provided in \ref{sec:frag-error}; here, we sketch the derivation and build on the result.

The fidelity between a system evolved using our fragmented procedure with that of the full system can be expressed as: 
\begin{equation}\label{eq:fid}
    F(t) = |\langle \Psi| U^{\dagger}(t) U_{I}^{(f)}(t) |\Psi \rangle|^2 .
\end{equation}
The unitary operator $U(t) = \exp(-iHt)$ evolves the system exactly under the full Hamiltonian $H$, while $U_{I}^{(f)}(t) = \exp(-iH_{I}^{(f)}t)$ evolves the system under a fragmented Hamiltonian that neglects interactions crossing the interface of fragment $f$. In \ref{sec:frag-error}, this expression is expanded for short times $t$ to understand how the evolution error $\epsilon(t) = 1 - F(t)$ depends on the strength of the neglected interactions. Through the use of Taylor expansion and the Baker–Campbell–Hausdorff (BCH) formula \cite{gilmore_bakercampbellhausdorff_1974}, we arrive at the first non-vanishing correction to the fidelity: 
\begin{equation}
    F(t) \approx 1 - \text{var}( H - H_{I}^{(f)}) t^2 ,
\end{equation}
where $\text{var}(\mathcal{O})$ is the quantum variance of operator $\mathcal{O}$. The error $\epsilon(t) = 1 - F(t)$ is thus given by $\text{var}( H - H_{I}^{(f)}) t^2$ for short times $t$.

The form of the short-time error provides a simple rule for choosing the target auxiliary qubits for fragment $f$ to minimize error; namely, select the environment qubit(s) whose interactions contribute most significantly to the variance $\text{var}(H - H_{I}^{(f)})$. This choice will minimize the short-time error of evolving the state by the fragmented Hamiltonian, which will lead to higher fidelity performance, on average (see Section \ref{sec:opt-aux}). Moreover, if the auxiliary selection is updated sufficiently often, the selection becomes exact as the short-time error dominates from the time of one auxiliary encoding to the next. 

\subsection{
Practical Implementation of the Optimal Auxiliary Encoding}\label{sec:using-aux}

Although the final form of the short-time evolution error provides insight into optimal auxiliary selection, the procedure for estimating a particular qubit’s contribution to the error within the distributed framework is less straightforward. For a general spin Hamiltonian, this variance is given by: 
\begin{equation} \label{eq:var}
    \begin{split}
    \text{var}(H - H_{I}^{(f)}) &= \text{var}\bigg( -\sum_{\langle i,j \rangle \in I} \sum_{\alpha, \beta} J_{ij}^{\alpha, \beta} \hat{S}^{(i)}_{\alpha} \hat{S}^{(j)}_{\beta}\bigg)
    \\&= \sum_{\langle i,j \rangle \in I} \sum_{\langle i',j' \rangle \in I} \sum_{\alpha, \beta} \sum_{\alpha', \beta'} J_{ij}^{\alpha, \beta} J_{i'j'}^{\alpha', \beta'} \big( \langle \hat{S}^{(i)}_{\alpha} \hat{S}^{(j)}_{\beta} \hat{S}^{(i')}_{\alpha'} \hat{S}^{(j')}_{\beta'}\rangle - \langle \hat{S}^{(i)}_{\alpha} \hat{S}^{(j)}_{\beta} \rangle \langle \hat{S}^{(i')}_{\alpha'} \hat{S}^{(j')}_{\beta'} \rangle \big) .
    \end{split}
\end{equation}
Estimating the full variance of Eq. (\ref{eq:var}) requires 4-point correlation measurements. If the distributed simulators are linked solely via classical channels, correlation measurements are only accessible when all relevant qubits are local to a single fragment. This implies that two auxiliary qubits -- one targeting $j$ and one targeting $j'$ -- must already be placed within the fragment in order to access the required 4-point correlator measurements. For $N_E$ environment qubits, there are $\mathcal{O}(N_{E}^{2})$ combinations, requiring $\mathcal{O}(N_{E}^{2})$ copies of the system in order to estimate all required 4-point correlators, undermining (although not necessarily precluding) the motivations for fragmented quantum simulation with such a technique.

The correlator calculation simplifies significantly when the variance is calculated with respect to a known product state, but a new issue arises: for many spin model Hamiltonians, the variance will vanish for certain initial product states. In fact, for the case of the transverse field Ising model \cite{stinchcombe_ising_1973}, this quantity vanishes for all computational basis states, providing no insight into the proper auxiliary choice. 

We propose a two-part solution that addresses these issues. First, we propose a proxy $v(a)$ that estimates the contribution of one potential auxiliary $a$ to the variance:
\begin{equation}
    v(a) = \sum_{\langle i,j \rangle \in I} \sum_{\langle i',j' \rangle \in I} \sum_{\alpha, \beta} \sum_{\alpha', \beta'} J_{ij}^{\alpha, \beta} J_{i'j'}^{\alpha', \beta'} \delta_{j,a} \delta_{j',a} \big( \langle \hat{S}^{(i)}_{\alpha}\hat{S}^{(j)}_{\beta} \hat{S}^{(i')}_{\alpha'} \hat{S}^{(j')}_{\beta'}\rangle - \langle \hat{S}^{(i)}_{\alpha} \hat{S}^{(j)}_{\beta} \rangle \langle \hat{S}^{(i')}_{\alpha'} \hat{S}^{(j')}_{\beta'} \rangle \big) .
\end{equation}
Inserting the two Dirac delta functions $\delta_{j,a} \delta_{j',a}$ eliminates the cross-terms in Eq. \ref{eq:var} that depend on multiple environment qubits. Thus, a single auxiliary is required to estimate $v(a)$, and in total $\mathcal{O}(N_{E})$ partitions are required to acquire $v(a)$ for all potential auxiliary targets. In addition to requiring fewer measurements, this proxy focuses on $a$'s contribution to the variance while neglecting the cross-terms that involve contributions from other potential auxiliary qubits. Secondly, to avoid scenarios where the variance vanishes for initial product states, we suggest first evolving the system for one time step $dt$ for a particular choice of $a$ before estimating $v(a)$. Although this procedure is more involved than calculating $v(a)$ for the initial product state directly, the overhead remains linear in the number of potential auxiliary qubits.

\section{Application 1: Fragmented Time Evolution} \label{sec:frag-TE}

\subsection{Simulators Linked via Classical Information} \label{sec:qc-TE}
We first focus on the scheme free of quantum information transfer, where the auxiliary qubits are selected at the beginning of the simulation and fixed to target a single environment qubit throughout the evolution.  We refer the reader to Fig. \ref{fig:link-comp} and \ref{sec:toy-model} for an in-depth look at how a system is fragmented for time evolution.
\begin{figure*}[ht]
    \centering
    \begin{minipage}[b]{2.5in}
    \begin{subfigure}[b]{2.5in}
        \centering
        \includegraphics[width=2.5in]{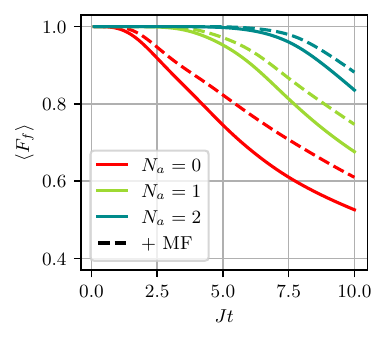}\captionsetup{justification=justified,singlelinecheck=false}
        \caption{}\label{subfig:evolution}
    \end{subfigure}
    \end{minipage}
    \begin{minipage}[b]{4.0in}
    \begin{subfigure}[b]{4.0in}
        \centering
        \includegraphics[width=4.0in]{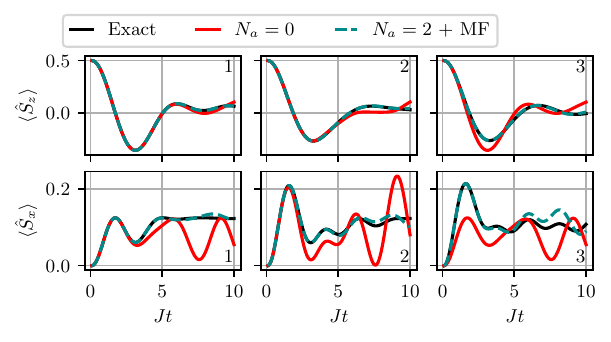}\captionsetup{justification=justified,singlelinecheck=false}
        \caption{}\label{subfig:expectation}
    \end{subfigure}
    \end{minipage}
    \caption{Nearest neighbor TFIM with constant $J = 1.0$. To produce \ref{subfig:evolution}, $N=12$ qubits are split into two fragments, and the fidelity between the fragment qubits' state and the exactly evolved system is plotted for various numbers of auxiliary qubits, with (dashed lines) and without (solid lines) mean-field corrections. The fidelity is averaged over non-zero $h$ values ranging between $\pm 1$. Performance progressively increases with increasing $N_a$ and the addition of mean-field corrections. \ref{subfig:expectation} displays the local expectation of $\hat{S}_z$ and $\hat{S}_x$ for a system of $N = 12$ qubits for the specific case of $h = 1.0$, with the corner label indicating site index. Here, we fragment the system into four fragments, each containing three qubits, and contrast the case of no communication (in red) to that of including $N_a = 2$ auxiliary qubits and mean-field corrections, which match the exact expectation values for longer simulation times.}
    \label{fig:fragment-evolution}
\end{figure*}
As a representative example, consider the transverse field Ising model (TFIM) \cite{pfeuty_one-dimensional_1970, stinchcombe_ising_1973} with a uniform transverse field: 
\begin{equation}
    H_{TFIM} = - J \sum_{\langle i,j \rangle} \hat{S}_{z}^{(i)} \hat{S}_{z}^{(j)} - h \sum_{i}^{N} \hat{S}_{x}^{(i)}.
\end{equation}
This system has been studied in depth to better understand the physics of quantum phase transitions \cite{sachdev_quantum_1999, dziarmaga_dynamics_2005, suzuki_quantum_2012}. 
We consider the evolution of a quantum system initialized in the computational basis state $|\mathbf{0}\rangle$ under $H_{TFIM}$, implementing exact unitary evolution numerically using PennyLane \cite{bergholm_pennylane_2022} with mean-field measurements updated every $dt = 0.1 / J$. Fig. \ref{fig:fragment-evolution} displays the scheme's performance for a 12-qubit model with nearest neighbor interactions of $J = 1.0$. The results presented in Fig. \ref{subfig:evolution} are averaged over non-zero transverse fields $h$ ranging from $\pm 1$, while Fig. \ref{subfig:expectation} features the specific case of $h = 1.0$. In Fig. \ref{subfig:evolution}, the system is split into two fragments, each simulating six of the system qubits and some number of auxiliary qubits. The average of the quantity $F_f$ is plotted, which we define as the fidelity between the reduced density operator of the system qubits within the fragment (tracing out any auxiliary qubits $a$) and the reduced density of the same system qubits for the exact evolution of the full system (tracing out all environment qubits forming $E$). We use the generalization of fidelity for density matrices \cite{liang_quantum_2019} to enable the focused evaluation of the fragment sub-system: 

\begin{equation}
    F_f = \bigg( \textrm{Tr} \textrm{ } \sqrt{\sqrt{\rho_f}\rho^{ex}_f\sqrt{\rho_f}} \bigg)^2 ,
\end{equation}

\begin{equation}
    \rho_f = \textrm{Tr}_a \textrm{ }\rho_{f+a} ,
\end{equation}

\begin{equation}
    \rho^{ex}_f = \textrm{Tr}_E \textrm{ }\rho^{ex} .
\end{equation}
For a short evolution time, the scheme captures the correct state of the system qubits within the fragment. This time can be extended by the inclusion of additional auxiliary qubits. 

In Fig. \ref{subfig:expectation}, we consider a specific instance of the TFIM with $J = 1.0$ and $h = 1.0$. To test the scheme, we split the system into smaller partitions with $N_f = 3$. When we make use of two auxiliary qubits and mean-field corrections, the local expectation values exhibit little error for several units of $Jt$, as expected from the fidelity results. 
%The qubits in the chain with site labels 4-6 belong to an ``interior'' fragment, whose interface consists of two interactions -- leading to larger errors in the corresponding expectation values.

\begin{figure*}[ht]
    \centering
    \includegraphics[width=5.0in]{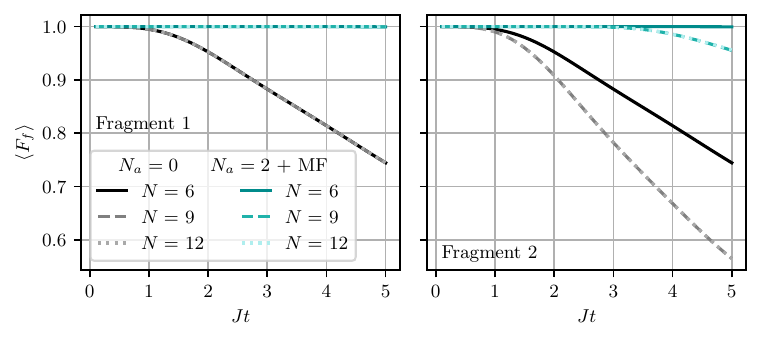}
    \caption{Scaling performance of the classical scheme for the nearest neighbor TFIM with $J = 1.0$, averaged over $h$ values ranging from $\pm 1$. The number of system qubits within a fragment $N_f$ is kept constant as $N$ is increased, with $N_a$ fixed to be zero (no communication, in black/gray) or two (in blue). The left panel plots $\langle F_f \rangle$ for the first fragment (which includes the boundary qubit and thus only involves one interaction crossing the interface), while the right panel plots $\langle F_f \rangle$ for the second fragment (which, for $N = 9$ and $N = 12$, is an \textit{interior} fragment with two interactions crossing the interface).}
    \label{fig:fragment-scaling}
\end{figure*}

In Fig. \ref{fig:fragment-scaling}, we examine the scaling performance for the nearest neighbor model, increasing the number of fragments simulated with increasing $N$ (keeping $N_f$ constant) with $N_a = 2$. In the left panel, the first fragment is considered. This fragment contains the boundary of the chain, and consequently, the fragment's interface consists of only one missing interaction. The right panel considers the second fragment, which is on the interior of the chain for $N > 6$ and consequently neglects two interactions, leading to reduced performance. This is manifested in the reduced fragment fidelity going from $N = 6$ to $N = 9$ for fragment 2. However, there is no such visible drop going from $N = 9$ to $N = 12$ due to the monogamy of entanglement \cite{de_oliveira_monogamy_2014} -- that is, although the number of qubits in the system grows, the qubits that are most strongly entangled with each other remain local to one fragment, and thus the amount of lost information shrinks as $N$ is further increased. We therefore expect our classical scheme to scale well with $N$ for systems that are locally interacting, and to serve as a strong approximation for moderate evolution times.

\subsection{The Addition of Quantum Information Transfer} \label{sec:qq-TE}

Next, we examine the case of selective quantum information transfer between quantum simulators, applicable when non-local operations are available, even if only in a limited capacity. In this case, the fragmentation scheme can be modified to include limited quantum information transfer (a \textit{quantum channel} \cite{cuomo_towards_2020}). The role of the auxiliary qubits shifts from being bystanders confined to a single fragment to qubits that are physically shared between simulators through selective non-local interactions, accomplished through qubit shuttling \cite{noiri_shuttling-based_2022} or teleportation \cite{bennett_teleporting_1993, bouwmeester_experimental_1997} (see Fig. 1c). If the simulations are being executed in parallel on a single quantum simulator \cite{das_case_2019, park_quantum_2022}, this would only require a few additional SWAP gates to include a limited number of cross-simulation interactions. 
In addition to providing more complete information transfer, a quantum channel further enables the active correction of auxiliary encoding as the system evolves. The selected number of auxiliary qubits places a limit on the number of qubits that are physically teleported / shuttled to a fragment; however, which environment qubits play this role can be changed from one time step to the next depending on which potential auxiliary qubit(s) have the largest contribution to the most recent estimate of the short-time error. When quantum channels and synchronized measurements are available, all correlation measurements are accessible. The quantity $v(a)$ can thus be estimated for any $a$ at any time. As the potential auxiliary qubits' contribution to the variance shift, new auxiliary qubits can be selected -- that is, we can make a new selection for which qubit(s) physically interact with a fragment native to a different simulator. If the time steps are sufficiently small such that the first non-vanishing order in the error dominates, then this becomes optimal even for long simulation times.  

To evaluate this scheme numerically, we abstract away the details of information transport; this topic has been investigated by other research in the context of distributed time evolution \cite{buessen_simulating_2023}. In our actively updated simulation, the quantity $v(a)$ is calculated for each potential auxiliary qubit at each time step to determine its contribution to the short-time error. This requires the estimation of correlators between each potential auxiliary and each fragment system qubit (requiring $\mathcal{O}(N_f N_E N_{\alpha\beta}^2)$ per fragment, where $N_{\alpha\beta}$ is the number of $\alpha\beta$ interaction types), but if there is only one kind of interaction (as is the case for the TFIM and other Ising-like models, with $\alpha = \beta = z$), all relevant correlators can be estimated from a set of full system snapshot measurements. The largest contributors are selected to be auxiliaries -- numerically, this amounts to keeping the interactions between these qubits and the fragment qubits, while zeroing the $J^{\alpha\beta}_{ij}$ coefficients of all other environment-fragment interactions (see \ref{sec:Jij}).
Any zeroed interactions can be approximately included via mean-field corrections. At the next time step, the selection of zeroed interactions might change due to a change in the selected auxiliaries for each fragment, as dictated by the short-time error. 

\begin{figure*}[ht]
    \centering
    \includegraphics[width=4.875in]{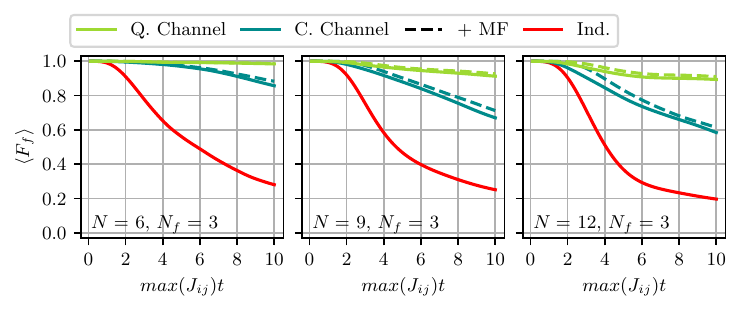}
    \caption{Comparison between scheme involving only classical information transfer (dark teal) to that involving limited quantum and classical information transfer (light green). For reference, the independent case (no information transfer) is included in red. The results are averaged over 100 Ising-like Hamiltonians with constant $h = 1.0$ and randomly generated graphs $J_{ij}$ (see Section \ref{sec:qq-TE}). The $N$ qubits are split into groups such that $N_f = 3$, with an additional $N_a = 2$ auxiliary qubits employed in the simulation.}
    \label{fig:qq-fidelity}
\end{figure*}

Fig. \ref{fig:qq-fidelity} compares this scheme (labeled ``Q. Channel'' to indicate the addition of quantum information transfer) to the previous scheme in Section \ref{sec:qc-TE}, which involves only classical information transfer (``C. Channel''). The graph plots the fragment fidelity $F_f$ averaged over 100 transverse field Ising-like models with $h = 1.0$ and randomly generated graphs $J_{ij}$. Each edge $ij$ exists with probability 0.5, and edge weights $J_{ij}$ are sampled from a Gaussian distribution with mean $\mu = 0.0$ and width $\sigma = 1.0$. Furthermore, we randomly select a computational basis state to initialize the fragmented system. Although both schemes outperform the case of no information transfer (in red), the complicated long-range nature of the Hamiltonians considered challenges the previous scheme, which only employs classical information transfer. In contrast, the quantum scheme preserves a large fragment fidelity, even at late simulation times. 

\subsection{Short-Time Error Auxiliary Selection}\label{sec:opt-aux}
\begin{figure*}[ht]
    \centering
    \includegraphics[width=6.5in]{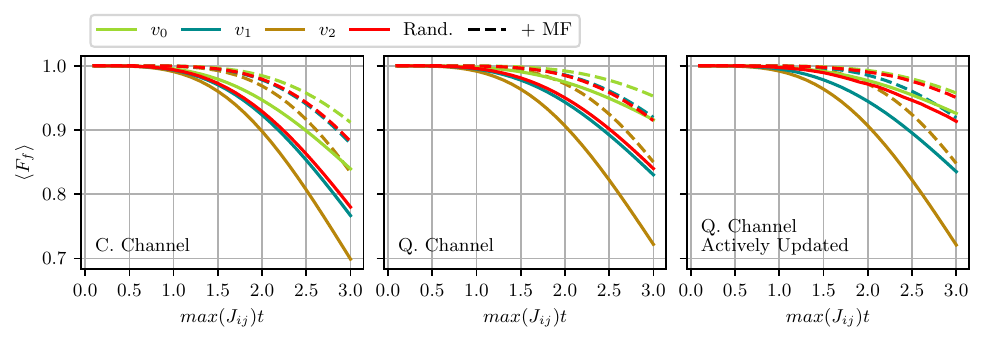}
    \caption{The effect of auxiliary selection on simulation performance. The curves are the averaged results for 100 different transverse field Ising-like Hamiltonians with $h = 1.0$ and the randomly generated graphs $J_{ij}$ described in Section \ref{sec:qq-TE}). Here, $N = 12$ with $N_f = 6$ and $N_a = 2$. The auxiliary target choices are ranked according to the size of $v(a)$, such that the two environment qubits with the largest $v(a)$ are used in simulation $v_0$, the two with the smallest $v(a)$ are used in simulation $v_2$, and the remaining two auxiliary choices are used in simulation $v_1$. Additionally, in red, we consider the case of randomly selecting two auxiliary target qubits with no variance calculation. In the left panel (classical communication) and center panel (quantum communication), the selection is made after one time step, and the choice remains fixed throughout evolution. In the right panel (quantum communication), the selection is re-evaluated at each time step.}
    \label{fig:aux-comp}
\end{figure*}

The benefit of using short-time error to inform auxiliary selection can be isolated by evaluating the performance of each auxiliary choice independently. Consider a system of $N = 12$ qubits, fragmented into two groups of $N_f = 6$. This leaves six environment qubits from the perspective of each fragment that could be targeted by an auxiliary qubit. Selecting two auxiliary qubits ($N_a = 2$), we rank the six potential choices for target auxiliary encoding according to the size of $v(a)$. In Fig. \ref{fig:aux-comp}, the six target encoding choices are divided into three groups of two based on $v(a)$, and each option is explored for randomly generated transverse field Ising-like Hamiltonians with $h = 1.0$, as considered in the previous section. A total of 100 such Hamiltonians are generated and simulated; the averaged results are presented in Fig. \ref{fig:aux-comp}, where $v_0$ corresponds to encoding the two environment qubits with the largest value for $v(a)$. On the left, the results are plotted for the case of classical information transfer. Any separation between the fidelity curves corresponding to different auxiliary choices indicates that the $v(a)$ metric meaningfully separates the potential auxiliary choices according to fidelity performance. The fact that the ordering corresponds to the ranked choice is evidence that using short-time error to select auxiliary encoding propagates to better performance at later times. In red, we consider random auxiliary encoding. The random performance roughly converges to the middle-ranked choice $v_1$ and can be thought of as the performance averaged over auxiliary encoding. In the center, the results are plotted for the case of additional quantum information transfer, without actively updating the auxiliary encoding. The results qualitatively match those of the classical case, with slightly better performance overall, consistent with Fig. \ref{fig:qq-fidelity}. In the right panel, we consider the quantum channel with actively updated auxiliary encoding. In this case, $v_0$ ($v_2$) corresponds to selecting the two auxiliary targets with the largest (smallest) values for $v(a)$ at each decision. The performance of $v_0$ marginally increases with the introduction of active updates, while the performance of $v_2$ marginally decreases. However, the random performance increases most markedly. Here, the rapid shuffling of auxiliary qubit encoding allows the fragments to quickly share information, leading to performance comparable to the optimal variance choice, $v_0$. The random, actively updated case has the added advantage of being measurement-efficient as it forgoes any variance estimation, but the highly frequent change of auxiliary encoding may lead to an overhead in qubit routing / swapping in order to be realized.

Finally, we note that in the averaged results presented in Fig. \ref{fig:aux-comp}, the mean-field corrected simulation (plotted with a dashed line) outperforms the corresponding simulation that fully neglects these interface interactions for every case considered. \ref{sec:benefit-of-mf} investigates the use of mean-field corrections to reduce simulation error through a numerical study. 

\section{Fragmented Quantum Circuits}\label{sec:frag-circ}
\begin{figure*}[ht]
    \centering
    \includegraphics[width=5.0in]{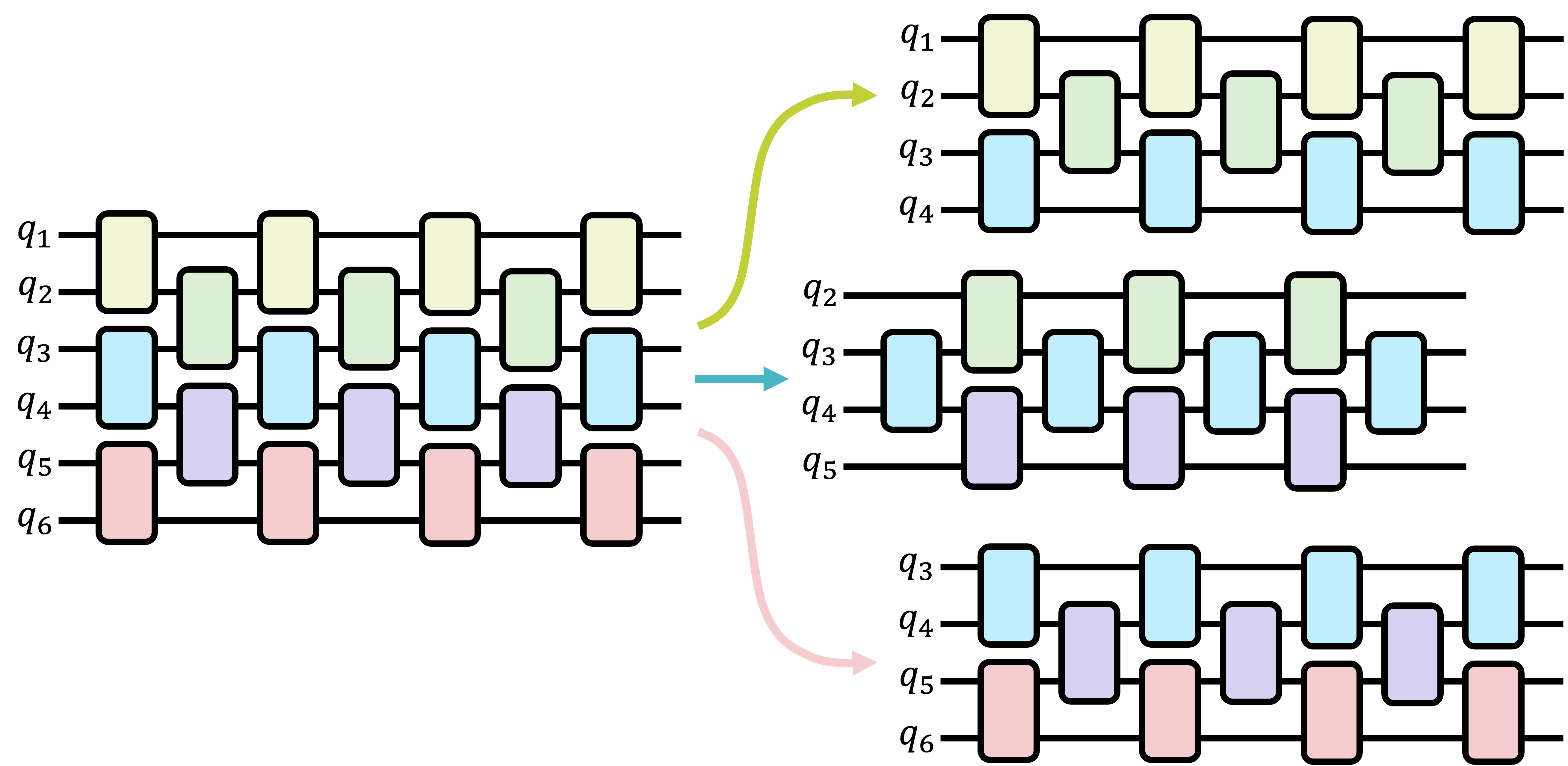}
    \caption{Diagram depicting how a circuit can be fragmented into a number of smaller circuits with overlapping registers, analogous to the inclusion of auxiliary qubits. The $N = 6$ qubits are partitioned into three groups of two (with $q_1, q_2$ addressed by the top PQC, $q_3, q_4$ addressed by the middle PQC, and $q_5, q_6$ addressed by the bottom PQC). Two additional auxiliary registers are included in each small PQC, such that some of the parameterized two-qubit gates appear in multiple PQCs. Gates that address qubits beyond the scope of one PQC are neglected by that particular circuit.}
    \label{fig:PQC-fragmented}
\end{figure*}

We now investigate the use of fragmentation in quantum circuit evolution. Consider the fragmentation of a parameterized quantum circuit (PQC) of size $N$ into multiple smaller PQCs. To fragment a circuit, multi-qubit unitaries that act on qubits outside the $N_{f+a}$ qubits devoted to a single sub-system's PQC are neglected. Although this resembles the first step of circuit cutting techniques \cite{peng_simulating_2020}, no data processing is required to reconstruct the cut gates; they are simply ignored. Crucially, some auxiliary qubits are included in each sub-system PQC, such that the full set of sub-system PQCs overlap with one another and $\sum_{f} N_{f+a} > N$ (see Fig. \ref{fig:PQC-fragmented}). The collection of fragmented circuits can be optimized alone prior to optimizing the full circuit as a new approach to \textit{pre-training}, commonly employed to boost variational quantum algorithms \cite{grant_initialization_2019,egger_warm-starting_2021,patti2022markov,qi_classical--quantum_2022, 
dborin_matrix_2022, li_adapting_2023,  niu_warm_2023}. Pre-training generally uses classical resources and can greatly increase the accuracy of a variational algorithm's solution, which is crucial for many applications such as reaching chemical accuracy for quantum chemistry problems \cite{peruzzo_variational_2014, mccaskey_quantum_2019, feniou_overlap-adapt-vqe_2023}. Our pre-training approach is motivated by the fact that the parameter solutions of the smaller circuits are expected to be smoothly connected to the parameter solutions of the full quantum circuit, as explored by \cite{mele_avoiding_2022}. We constrain the pre-training to use small circuits that are cheap to simulate classically. Furthermore, employing smaller circuits limits entanglement growth, which has been shown to improve training and avoid barren plateaus \cite{mcclean_barren_2018, holmes_barren_2021, ortiz_marrero_entanglement-induced_2021, patti_variational_2022, mele_avoiding_2022}.

\section{Application 2: Fragment-Initialized VQE} \label{sec:frag-VQE}
Our method of fragmenting a quantum circuit can be applied to classically pre-train quantum circuit parameters for the variational quantum eigensolver (VQE) \cite{fedorov_vqe_2022}. For this application, a PQC of size $N$ is divided into smaller PQCs, each having size $N_{f+a} < N$. To optimize each sub-system PQC, the mean-field-corrected Hamiltonian given in Eq. (\ref{eq:Ham-MF}) is minimized. In addition to facilitating the study of quantum systems and statistical physics, mean-field methods have been introduced for data analysis and loss function modification \cite{opper_mean_1996, hojen-sorensen_mean-field_2002, gao_mean_2003, patti_multibasis_2022}. In our pre-training technique, employing mean-field terms serves to link the optimization of the separate circuits by their current mean-field measurements. Overlapping parameters (that is, parameters shared by two fragmented PQCs) are initialized for one PQC using the most recent values from the other, further uniting the separate circuit optimizations. The mean-field measurements are updated regularly, and optimization halts when the steady state (up to some set precision) is reached for all parameters -- those shared and those unique to one PQC -- or the maximum number of iterations is reached. The algorithm is outlined in Algorithm \ref{alg:MF-VQE}.

\begin{minipage}{5.0in}
  \begin{algorithm}[H]
    \caption{Fragment pre-training with mean-field corrections.}\label{alg:MF-VQE}
    \begin{algorithmic}
    \State (Randomly) initialize $\{\theta_i\}$ for the brickwork section of the full PQC.
    \State Divide $\{\theta_i\}$ into a set $\{\theta_{f,i}\}$ for each fragment $f$. 
    \State Initialize $\langle \hat{S}^{(j)}_\beta \rangle(0) = 0$. 
    \Repeat 
    \For{$f$ in system}
    %\State $\theta_{f, i=a}(k) \leftarrow \theta_{i=a}(k)$ for auxiliary spins $a$.
    \State $\theta_{f, i=a}(k) \leftarrow \theta_{i=a}(k)$ for auxiliary spins $a$ in $f$.
    \State $\theta_{f, i}(k+1) \leftarrow \theta_{f, i}(k) - \eta \nabla_{\theta_{f, i}} \langle H_{MF}^{(f)} \rangle_f$.
    \State $\langle \hat{S}^{(j)}_\beta \rangle(k+1) \leftarrow \langle \hat{S}^{(j)}_\beta \rangle_f(k+1) $ for system spins $j \in f$.
    \State $\theta_{j}(k+1) \leftarrow \theta_{f, j}(k+1)$ for system spins $j \in f$. 
    \EndFor
    \Until{Parameters $\{\theta_{i}\}$ converge.}
    \end{algorithmic}
    \end{algorithm}
\end{minipage}
\\

\subsection{Details of Ansatz}
\begin{figure*}[ht]
    \centering
    \includegraphics[width=5.0in]{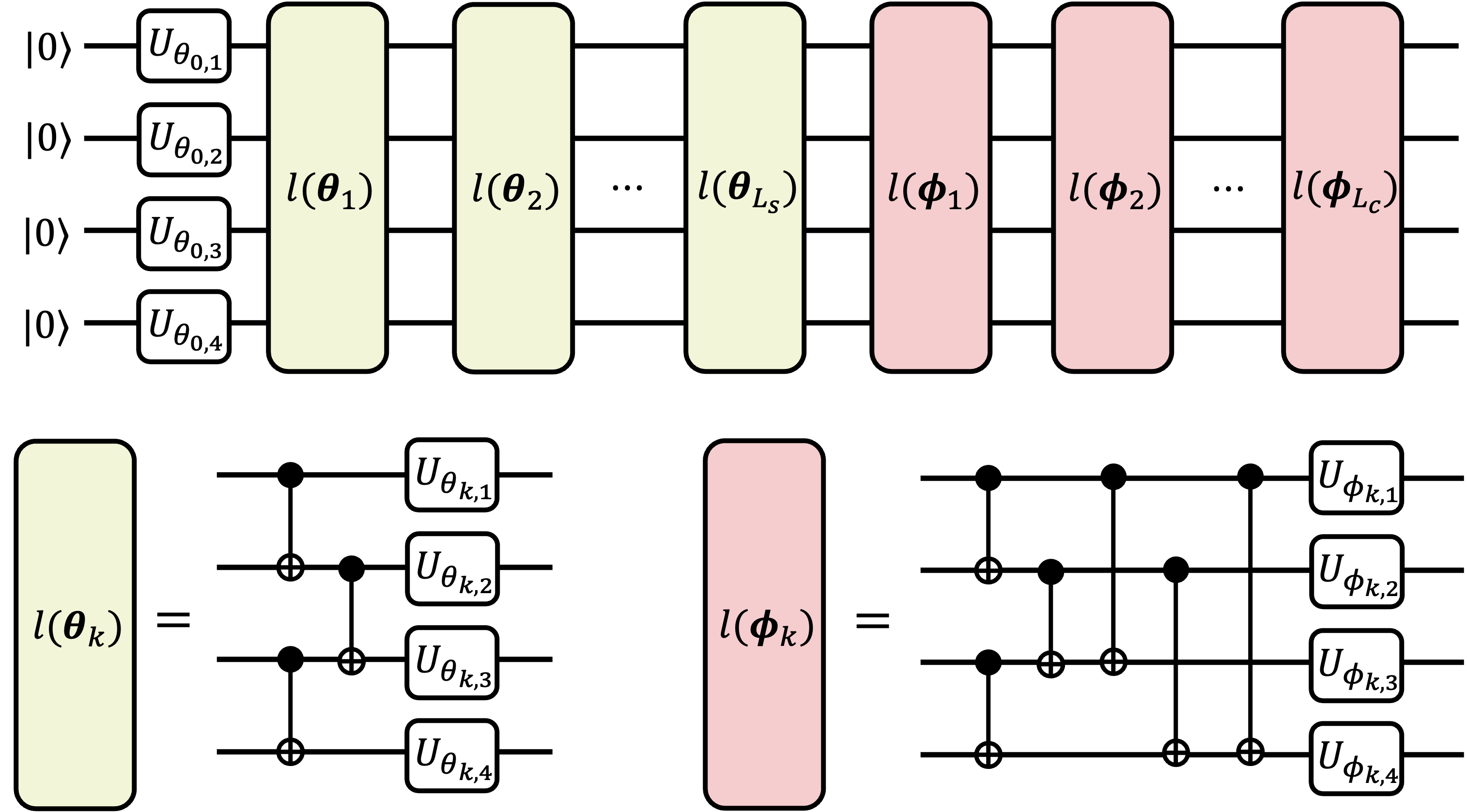}
    \caption{The circuit ansatz is built from $L_s$ layers $l(\theta)$ with linear entangling gates, which are amenable to fragmentation. These are followed by a set of $L_C$ layers $l(\phi)$ with an all-to-all entangling architecture. Only the brickwork layers parameterized by $\theta_i$ are pre-trained using the fragmented scheme, while the layers parameterized by $\phi_j$ are employed only in the final training process.}
    \label{fig:PQC}
\end{figure*}

We focus on pre-training brickwork circuits with a limited number of layers to constrain entanglement growth between fragments. Although a circuit ansatz with high complexity is often necessary for interesting VQE applications in order to provide enough expressivity to reach the ground state \cite{benedetti_parameterized_2019, du_expressive_2020}, fragmentation-based pre-training is still beneficial through the use of a layer-wise approach \cite{liu_layer_2022}. If a shallow brickwork circuit is placed ahead of a more expressive PQC ansatz, the brickwork layers can first be optimized using the fragmented approach. These layers serve to bring the state of the system to have some ground state overlap. The full circuit VQE can then be performed, initializing the leading brickwork layers of the circuit with the pre-trained parameter values and initializing the remaining gates of the ansatz to approximately act as identity -- specifically, we choose to randomly initialize these parameters to be small values bounded by $\pm \varepsilon$ (with $\varepsilon = 10^{-5}$ for our results), to balance maintaining the optimized action of the initial layers after pre-training while avoiding training issues associated with a true identity initialization \cite{grant_initialization_2019, zhao_zero_2022}. The overall circuit layout is outlined in Fig. \ref{fig:PQC}.

\subsection{Performance and Analysis of Fragmented Pre-Training}
To evaluate our pre-training method, we focus on random Ising-like models. In Section \ref{sec:maxcut_results}, we present the numerical performance of the scheme for the classical case of zero transverse field ($h = 0$). Having established the advantage of the approach, in Section \ref{sec:solve-maxcut} we derive its success as stemming from the mean-field corrective terms included in the loss function, which shift the global minimum of the collective fragmented circuit to coincide with that of the full optimization problem. Finally, in Section \ref{sec:mf-pt}, we use perturbation theory and numerical simulation to demonstrate that our approach remains beneficial for $|h| > 0$, in the regime of a weak transverse field.

\subsubsection{VQE Results for MaxCut}\label{sec:maxcut_results}
We first benchmark the scheme using randomly generated classical Ising Hamiltonians, where the all-to-all $J_{ij}$ interactions are sampled from a Gaussian distribution (mean $\mu = 0.0$, width $\sigma = 1.0$) and the transverse field $h$ is fixed to be zero. These models can be mapped to MaxCut problems with randomly generated graphs \cite{patti_variational_2022}. For the circuit ansatz, a fixed number of brickwork layers is used ($L_s = 4$) to keep this portion of the circuit shallow, while the all-to-all entangling portion of the circuit is made up of $L_c = N$ layers. The parameterized single qubit rotations within each layer are selected to be one rotation about $x$ followed by one rotation about $y$, and the entangled gates are selected to be controlled $z$ (CZ) rotations. All simulations are performed numerically using PennyLane \cite{bergholm_pennylane_2022}. Lastly, note that parameter convergence (evaluated every 100 iterations) is used as the stopping criterion for both the fragmented circuit and full circuit optimization, with a maximum of 5000 iterations permitted. 

\begin{figure*}[ht]
    \centering
    \includegraphics[width=6.5in]{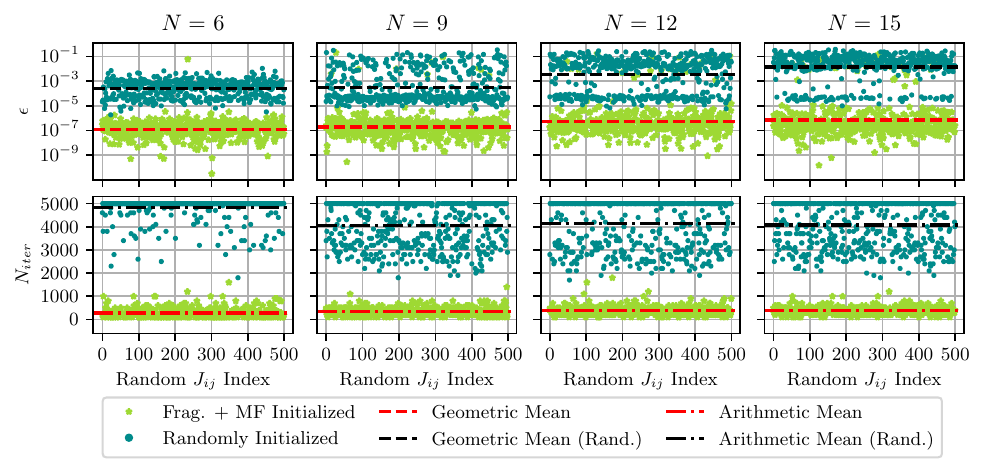}
    \caption{Comparison between fragment pre-trained VQE and vanilla VQE for 500 different $J_{ij}$ matrices (graphs). The full PQC is split into fragments with $N_a = 2$ and at most $N_f = 3$ during pre-training. A total of $T = 10$ different partitionings are considered, and the best pre-trained solution is used to initialize the final optimization. The final percent error $\epsilon$ is provided in the top panel, while the required number of iterations $N_{iter}$ to reach convergence is provided in the bottom panel. For the fragment initialized case, these metrics refer to the full circuit training that occurs after pre-training. Fragment pre-training reduces the geometric mean of $\epsilon$ by orders of magnitude, even as the system size increases. Likewise, the mean number of required iterations is reduced by nearly an order of magnitude.}
    \label{fig:VQE-data}
\end{figure*}

To assess the performance of pre-training using circuit fragmentation, the same circuit is optimized using random initial values (referred to as ``vanilla VQE''). Fig. \ref{fig:VQE-data} provides a case-by-case comparison between fragment-initialized VQE and vanilla VQE for 500 such models, for circuits of up to 15 qubits. In the top panels, the final percent error $\epsilon = (E - E_0) / |E_0|$ (where $E_0$ is the true ground state energy) is plotted for both approaches, along with the geometric mean of the results. The geometric mean of the fragment-initialized final error lies roughly three orders of magnitude below that of the vanilla VQE, with this gap growing even larger with increasing system size. For the larger system sizes, the vanilla VQE struggles to find a solution having $\epsilon < 10^{-2}$, while the fragment-initialized approach reaches $\epsilon \sim 10^{-7}$ for the same problem Hamiltonian. Moreover, using the same stopping criterion, the fragment-initialized VQE reaches this solution in fewer iterations ($N_{iter}$), decreasing the average number by nearly an order of magnitude, as illustrated by the bottom panels of Fig. \ref{fig:VQE-data}. After successful pre-training, the parameters of the stitched-together circuit produce a loss that is already in the neighborhood of the minimum, so fewer iterations are required to reach convergence. For this simulation, we employ a batched optimization of $T$ different fragmented circuits performed in parallel. See \ref{sec:batch-size} for a description of this approach.

\subsubsection{Solving MaxCut with Mean-Field Terms}\label{sec:solve-maxcut}
Our modification of fragmented loss functions to replace missing (that is, inaccessible) interactions with mean-field terms is critical to the success of pre-training. We here demonstrate that when there is no transverse field (as is the case for Ising-like Hamiltonians that map to classical graph problems), mean-field replacement of interactions results in a ground state and ground state energy that coincide with that of the exact Hamiltonian. This can be shown using a simple logical argument. First, it is well-established that the ground state of a classical Ising Hamiltonian will be a computational basis state -- indeed, this is why the ground state can be mapped to the solution of a classical problem. We denote the ground state by $|x^* \rangle$. The ground state energy is simply a sum of the expected values of weighted $ZZ$ interactions, taken with respect to the computational basis state $|x^*\rangle$: $E_{g} = -\sum_{\langle i, j\rangle} J_{ij} \langle x^* | \hat{S}_{z}^{(i)}\hat{S}_{z}^{(j)}| x^* \rangle$. Notice that for any computational basis state $|x\rangle$, the value of the expectation of a $ZZ$ interaction exactly equals the value of the product of the expectation of the individual $Z$ operators; that is, $\langle x| \hat{S}_{z}^{(i)}\hat{S}_{z}^{(j)}|x \rangle = \langle x| \hat{S}_{z}^{(i)} | x \rangle \langle x | \hat{S}_{z}^{(j)}|x \rangle$. Thus, if any weighted interaction $J_{ij}\langle\hat{S}_{z}^{(i)}\hat{S}_{z}^{(j)}\rangle$ is replaced by its mean-field counterpart $J_{ij} \langle\hat{S}_{z}^{(i)}\rangle \langle \hat{S}_{z}^{(j)}\rangle$, the resultant energy is unchanged: $E_{g} = \langle x^* | H | x^* \rangle = \langle x^* | H_{MF} (|x^* \rangle) | x^* \rangle $, where $H_{MF}$ is the union of the fragmented, mean-field corrected Hamiltonians $\{H_{MF}^{(f)}\}$ and we have explicitly included the state dependence due to the presence of mean-field terms. Having established this fact, we must now show that $|x^*\rangle$ is the ground state of $H_{MF}$, such that $\langle x^* | H_{MF} (|x^*\rangle) | x^* \rangle \leq \langle \psi | H_{MF} (|\psi\rangle) | \psi \rangle \; \forall \; |\psi\rangle$. Observe that the quantity $\langle \hat{S}_{z}^{(i)}\hat{S}_{z}^{(j)} \rangle$ is bounded by $\pm 1/4$ and equals one of these extremum values for any computational basis state. The mean-field counterpart $\langle \hat{S}_{z}^{(i)} \rangle \langle \hat{S}_{z}^{(j)} \rangle$ shares the same bounds; therefore, we cannot expect any state $|\psi\rangle$ to produce a smaller energy $\langle \psi | H_{MF} (|\psi\rangle) | \psi \rangle$ than $|x^*\rangle$, the ground state of the full Hamiltonian. 

The above is a central reason for the success of our fragmented training: for Ising-like models with zero transverse field, optimizing a Hamiltonian with mean-field corrections will solve the original problem mapped to the full Hamiltonian. Two potential error sources can arise: 1) the state produced by stitching the optimized circuits together can differ from the output of the individual circuits, and 2) the fragmented optimization may have limited success, e.g., by landing in a local minimum or stalling in a barren plateau. A balance should be struck between these complications: the first error source can be mitigated by considering larger fragments with a larger number of auxiliary qubits or possibly by limiting the number of inter-fragment unitaries, as done in \cite{khait_variational_2023}, while the second can be mitigated by considering smaller fragments with fewer circuit parameters.

\subsubsection{Mean-Field Terms as First-Order Perturbation Corrections}\label{sec:mf-pt}
We now use perturbation theory to elucidate our technique of replacing multi-qubit interactions with mean fields when $h \neq 0$.
In the previous section, it is established that the ground state and ground state energy of an Ising-like Hamiltonian with zero transverse fields remain unchanged when one or more of the interactions are replaced by the corresponding mean-field approximation term. Following a similar argument, one can further establish that the computational basis states are stationary states of the mean-field corrected Hamiltonian $H_{MF}(|\psi\rangle)$, and therefore $H_{MF}(|\psi\rangle)$ and the unaltered Hamiltonian $H$ share the same spectrum and set of eigenstates (although this term is used loosely for $H_{MF}(|\psi\rangle)$, as the dependence on $|\psi\rangle$ causes the stationary Schrödinger equation to deviate from a linear eigenvalue problem). 

In this section, we consider adding a small transverse field to the classical Ising-like model, propelling the problem into the quantum domain. The first-order corrections to the ground state $|x^* \rangle$ and ground state energy $E_{g}$ are computed using perturbation theory. The case of the mean-field corrected Hamiltonian $H_{MF}(|\psi\rangle)$ is treated with a version of perturbation theory modified to accommodate mean-field terms, and notably, the same first-order corrections to $|x^* \rangle$ and $E_{g}$ are recovered. For a full derivation, please refer to \ref{sec:perturbation-theory}. 

Adding a transverse field, the unaltered Hamiltonian containing all interactions is given by: 
\begin{equation}
    H = H_{0} + H_{I} + \lambda V,
    \label{eq:H_perturbation}
\end{equation}
where $H_0$ contains the intra-fragment interactions:
\begin{equation}
    H_{0} = -\sum_{\langle i, j \rangle \notin I} J_{ij} \hat{S}_{z}^{(i)} \hat{S}_{z}^{(j)} ,
\end{equation}
$H_I$ contains the inter-fragment interactions:
\begin{equation}
    H_{I} = -\sum_{\langle i, j \rangle \in I} J_{ij} \hat{S}_{z}^{(i)} \hat{S}_{z}^{(j)} , 
\end{equation}
$V$ contains the perturbing transverse field:
\begin{equation}
    V = -h \sum_{i} \hat{S}_{x}^{(i)} ,
\end{equation}
and $\lambda$ is a perturbation parameter. We remind the reader that the inter-fragment interactions $H_{I}$ are those that will be replaced by mean-field corrections. 

In contrast, the mean-field corrected Hamiltonian denoted $H_{MF}$ is given by: 
\begin{equation}\label{eq:H_MF}
    H_{MF}(|\psi\rangle) = H_{0} + H_{I, MF}(|\psi\rangle) + \lambda V , 
\end{equation}
where the form of the Hamiltonian now depends on the state of the system due to the mean-field corrections:
\begin{equation}
    H_{I, MF}(|\psi\rangle) = -\sum_{\langle i, j \rangle \in I} J_{ij} \hat{S}_{z}^{(i)} \langle \psi | \hat{S}_{z}^{(j)} | \psi \rangle .
\end{equation}

Before any corrections can be computed, it is imperative to establish the correct zeroth order energies and eigenstates for each Hamiltonian. Following perturbation theory, the zeroth order eigenstates of $H$ and $H_{MF}$ generally equal those of the unperturbed counterparts (that is, taking $h = 0$); these coincide with the set of computational basis states $ \{ |x\rangle \}$ -- including the unperturbed ground state, $|x^* \rangle$. However, there are degeneracies in the unperturbed Hamiltonians, and thus, degenerate perturbation theory is required. 

When the unperturbed spectrum contains degeneracies, the proper linear combinations of the unperturbed eigenstates forming the degenerate subspace must be determined; these are the states that the perturbed eigenstates approach as $h \rightarrow 0$. The unperturbed Ising-like model possesses $\mathbb{Z}_2$ symmetry. Practically, this means that for each eigenstate $|x\rangle$, the ``flipped'' eigenstate $|\bar{x}\rangle := \bigotimes_i X_i |x_i\rangle$ is degenerate. For the unaltered Ising-like model $H$, the proper zeroth order eigenstates for the degenerate subspace containing the ground state are given by $|\pm_{x^{*}}\rangle = \frac{1}{\sqrt{2}} (|x^{*}\rangle \pm |\bar{x}^*\rangle)$. The transverse field will break the ground state degeneracy of $H$, and the positive superposition $|+_{x^{*}}\rangle$ is preferred by the ground state. 
  
Shifting attention to the mean-field corrected Hamiltonian $H_{MF}$, the stationary Schrödinger equation is no longer linear in $|\psi\rangle$, and the linearity that characterizes quantum mechanics no longer applies. The notion of finding proper linear combinations is not an appropriate procedure due to the problem's nonlinearity. In particular, superpositions of degenerate eigenstates can yield different energies for $H_{MF}$ and thus effectively exist outside the degenerate subspace. 

To illustrate this, consider a single mean-field factor, $\langle\hat{S}_{z}^{(i)}\rangle$, such as those within $H_{MF}$. While the expectation value of this quantity with respect to a computational basis state $|x^*\rangle$ yields 
\begin{equation}
    \langle x^* | \hat{S}_{z}^{(i)} | x^* \rangle = \frac{1}{2} (-1)^{x^*_{i}} ,
\end{equation}
evaluating the same term with respect to $|+_{x^*}\rangle$ leads to the term vanishing as
\begin{equation}
\begin{aligned}
    \langle +_{x^*} | \hat{S}_{z}^{(i)} | +_{x^*} \rangle &= \frac{1}{2} \big( \langle x^* | \hat{S}_{z}^{(i)} | x^* \rangle + \langle x^* | \hat{S}_{z}^{(i)} | \bar{x}^* \rangle + \langle \bar{x}^* | \hat{S}_{z}^{(i)} | x^* \rangle + \langle \bar{x}^* | \hat{S}_{z}^{(i)} | \bar{x}^* \rangle\big) 
    \\ &= \frac{1}{4} \big( (-1)^{x^*_{i}} + (-1)^{\bar{x}^*_{i}} \big)
    \\ &= 0 .
\end{aligned}
\end{equation}
Notably for the ground state of the unperturbed Hamiltonian $|x^*\rangle$, this means that the pure computational basis states $|x^*\rangle, |\bar{x}^*\rangle$ are energetically preferred over any linear combination of them. Thus, for $H_{MF}$, the computational basis states remain the proper zeroth order eigenstates with a perturbative transverse field. 

After establishing the zeroth order eigenstates and eigenenergies ($|k^{(0)}\rangle$ and $E_{k}^{(0)}$, respectively) of conventional Hamiltonians such as Eq.\ \ref{eq:H_perturbation}, perturbation theory proceeds by expanding $|k\rangle$ and $E$ in $\lambda$ in the stationary Schrödinger equation and equating orders of $\lambda$: 
\begin{equation}
\begin{aligned}
    \Big( H_{0} &+ H_{I} + \lambda V \Big) \big( 
    |k^{(0)}\rangle + \lambda|k^{(1)}\rangle + \lambda^2 |k^{(2)}\rangle + \cdots \big) 
    \\ &= \big( E_{k}^{(0)} + \lambda E_{k}^{(1)} + \lambda^2 E_{k}^{(2)} + \cdots\big)  \big( |k^{(0)}\rangle + \lambda|k^{(1)}\rangle + \lambda^2 |k^{(2)}\rangle + \cdots \big)  .
\end{aligned}
\end{equation}
Following this procedure for $H$ and carefully treating the degeneracy, the first-order energy correction $E_{k}^{(1)}$ vanishes and the first-order eigenstate correction takes the form: 
\begin{equation}\label{eq:pt-1st-order-psi}
    | k^{(1)} \rangle = \sum_{m \notin D_k} \frac{\langle m^{(0)} |V| k^{(0)} \rangle}{E_{k}^{(0)} - E_{m}^{(0)}} |m^{(0)} \rangle , 
\end{equation}
where $D_{k}$ represents the degenerate subspace that $|k^{(0)}\rangle$ occupies. 

To derive the analogous correction to $H_{MF}$, we employ a modified approach to perturbation theory that can accommodate the nonlinearity of the stationary Schrödinger equation. In particular, the expanded form of $|k_{MF}\rangle$ is explicitly inserted into the state-dependant terms of $H_{MF}$ prior to equating orders of $\lambda$ to compute the corrections. Following this procedure, the first order energy correction $E_{k, MF}^{(1)}$ again vanishes, and the first order eigenstate correction takes on an identical form to that of $H$: 
\begin{equation}\label{eq:mfpt-1st-order-psi}
    | k_{MF}^{(1)} \rangle = \sum_{m \notin D_k} \frac{\langle m_{MF}^{(0)} |V| k_{MF}^{(0)} \rangle}{E_{k, MF}^{(0)} - E_{m, MF}^{(0)}} |m_{MF}^{(0)} \rangle .
\end{equation}
There is one crucial difference between Eq. (\ref{eq:pt-1st-order-psi}) and Eq. (\ref{eq:mfpt-1st-order-psi}): the zeroth order eigenstates $|m^{(0)} \rangle$ and $|m_{MF}^{(0)} \rangle$. For the full Hamiltonian, each $|m^{(0)} \rangle$ has the form $|\pm_{y} \rangle \propto (|y\rangle \pm |\bar{y}\rangle)$, while each $|m_{MF}^{(0)} \rangle$ is a single computational basis state, $|y\rangle$. This leads to the fidelity between the first order ground state $|\psi_{g, MF} \rangle \propto |x^{*}\rangle + | x^{*(1)}_{MF}\rangle$ and that of the full Hamiltonian $|\psi_{g} \rangle \propto |+_{x^{*}}\rangle + |+_{x^{*}}^{(1)}\rangle$ to be $F = 0.5$ rather than perfect unity (see \ref{sec:overlap}). Nonetheless, half overlap provides significant information about the true ground state for pre-training. 
\begin{figure*}[ht]
    \centering
    \includegraphics[width=6.5in]{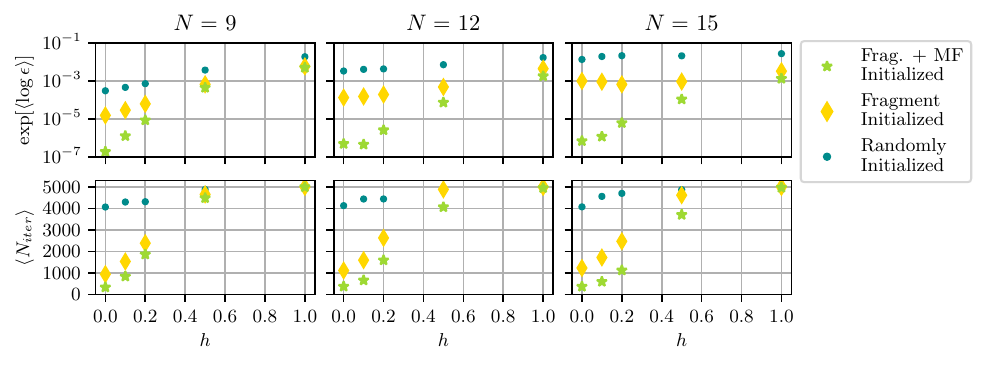}
    \caption{Comparison between fragment pre-trained VQE (with and without mean-field corrections) and vanilla VQE performance as a function of the transverse field strength $h$. Each point is the mean (geometric or arithmetic) performance of 500 different $J_{ij}$ matrices (graphs). The full PQC is split into fragments with $N_a = 2$ and at most $N_f = 3$ during pre-training. A total of $T = 10$ different partitionings are considered, and the best pre-trained solution is used to initialize the final optimization. The mean final percent error $\epsilon$ is provided in the top panel, while the mean required number of iterations $N_{iter}$ to reach convergence is provided in the bottom panel.}
    \label{fig:VQE-data-vs-h}
\end{figure*}

In Fig. \ref{fig:VQE-data-vs-h}, we examine the mean performance of the pre-training scheme as a function of the transverse field strength $h$. The case of neglecting mean-field terms during pre-training is also considered to highlight the vital role these corrections play at small values of $h$. When $h = 0$, the model is classical, and the global fragmented Hamiltonian with mean-field corrections shares the same ground state and ground state energy of the full model, as discussed in \ref{sec:solve-maxcut}. This leads to remarkable pre-training performance, even as $N$ is increased. The fragment initialization that neglects mean-field terms is not guaranteed to share the ground state of the full model -- in fact, the two are likely to be orthogonal. The pre-training will still feature low entanglement, which likely explains why the mean-field free initialization scheme outperforms random initialization, but overall, the average error exceeds that of the mean-field corrected case by orders of magnitude, particularly as $N$ is increased. When a small transverse field is added to the model, the half overlap between the fragmented and full Hamiltonian ground states provided by including mean-field corrections leads to error orders of magnitude smaller than the other approaches. Only as $h$ is further increased -- entering the regime where first-order perturbation theory is inadequate to describe the ground state -- do the approaches begin to perform comparably, with increasing error and required iterations.

\section{Conclusion} \label{sec:conclusion}
We have presented two near-term approaches to the distributed Hamiltonian evolution of a quantum system and a pre-training technique for variational quantum circuits. Our time evolution schemes are built upon the idea that the relative importance of interactions spanning a sub-system and its environment can be ascertained using the principle of minimizing the short-time evolution error, which is derived to be proportional to the quantum variance of the difference between the full and fragmented Hamiltonians. The first scheme employs only classical information transfer in the form of mean-field measurements to update mean-field corrections, as well as a limited number of auxiliary qubits anchored to each fragment, enabling limited entanglement growth. Although our method is lossy, metrics local to the system qubits addressed by a single fragment can closely mimic the true values from exact evolution, including the gold standard comparison of state fidelity. Moreover, this scheme is flexible, as it is amenable to any mixture of classical and quantum hardware and can process the fragments in series or parallel. In our second scheme, the information stored by qubits designated to be auxiliaries is physically shared between fragments, either through qubit shuttling or quantum teleportation. This approach is appropriate when quantum hardware is available and limited quantum communication is feasible. If desired, the choice of which qubits act as auxiliaries can be updated from one time step to the next, as dictated by the minimum error rule, to extend the performance of the approximate scheme.  

Finally, we examine how our fragmented simulation scheme can be modified to apply to quantum circuits. Here, a single circuit is fractured into several smaller overlapping circuits, which are more manageable (requiring lower connectivity and less prone to suffer from noise and barren plateaus) and, if sufficiently small, even classically treatable. We devise a scheme that employs fragmented circuits to pre-train the parameters of the full PQC. Crucially, the use of overlapping registers coupled with the mean-field corrective terms in the loss function links the optimization of the individual circuits. The inclusion of mean-field corrections shifts the solution of the collective circuit optimization to have a large overlap with the solution of the full problem. We demonstrate that the pre-training scheme reduces the final percent error by orders of magnitude as well as the number of iterations required when compared to randomly initialized full circuit optimizations of VQE. Although the scheme's performance is particularly strong for classical Ising Hamiltonians, we develop a non-linear perturbation theory to analytically show that the mean-field terms included in optimization act as first-order perturbation corrections when a small transverse field added, extending the success of the scheme into the quantum realm. 

This manuscript motivates and facilitates numerous future research directions. Although we emphasized limited quantum information transfer, subsequent studies might explore how the number of auxiliary qubits and the frequency of re-encoding affect distributed simulation, or devise the details of physically implementing a limited quantum channel. Likewise, higher-order moments beyond mean-field terms may be explored as higher-order corrections. Moreover, rather than using auxiliary qubits to target specific environment qubits, a method of mapping salient environment \textit{states} to auxiliary qubits (as employed by some classical fragmentation methods such as DMET \cite{wouters_practical_2016}) may further improve the method, although the measurement-efficiency of such a technique may prove challenging in a quantum setting. Regarding fragment pre-training for variational algorithms, future works might develop efficient circuit fragmentations that are tailored to specific problem Hamiltonians and/or symmetries, rather than our more general, batched approach. Finally, alternative partitioning schemes might be considered to enable pre-training of non-brickwork circuits.

This work represents a pivotal stepping stone on the path to large-scale distributed quantum computing. In the near term, our distributed computing method with classical channels can be implemented by a single small simulator in sequence, or by a collection of small simulators that are either quantum or classical in nature. This permits the simulation of large system quantum dynamics without the noise and connectivity concerns of a large-scale quantum device \cite{cheng_noisy_2023}, allowing experimentalists to address challenging problems in quantum chemistry and condensed matter physics \cite{smith_simulating_2019, kassal_polynomial-time_2008}. As non-local operations on quantum hardware improve, our proposal for limited quantum information transfer can be implemented, enabling cross-simulator measurements and higher accuracy. Lastly, our fragmented pre-training method can reduce the error of large-scale variational quantum algorithms by orders of magnitude while reducing the number of training epochs. Such improvements are vital to this field, which seeks to address problems ranging from drug discovery to NP-hard optimization on quantum hardware despite persistent training difficulties \cite{peruzzo_variational_2014, pyrkov_quantum_2023, salehi_unconstrained_2022, fedorov_vqe_2022, cerezo_variational_2021}.

\section{Acknowledgements}
This work was done during A.M.G.'s internship at NVIDIA. A.M.G. acknowledges support from the National Science Foundation (NSF) through the Graduate Research Fellowships Program, as well as support through the Theodore H. Ashford Fellowships in the Sciences. At CalTech, A.A. is supported in part by the Bren-endowed chair. S.F.Y. thanks the AFOSR and the NSF (through the CUA PFC and QSense QLCI) for funding.

\newpage
\bibliographystyle{unsrt}
\bibliography{main}

\newpage
\appendix

% NOTE: I've removed what was the first appendix (4th order correction to reorder qubits), since it didn't make a difference in the results (and it isn't being used in the newest results). 

\section{Fragmentation for Time Evolution}
\subsection{A Toy Model}\label{sec:toy-model}
\begin{figure*}[ht]
    \centering
    \includegraphics[width=6.5in]{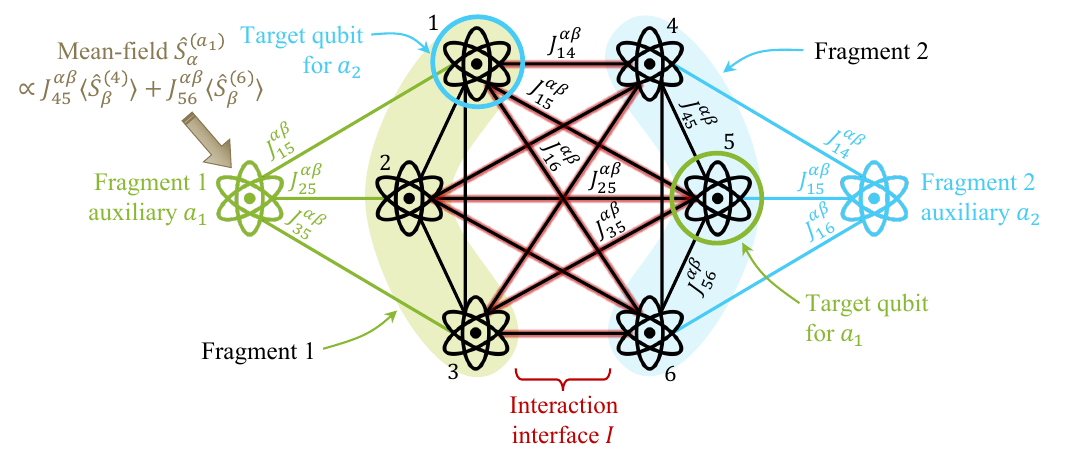}
    \caption{A schematic illustrating how a system of $N = 6$ qubits can be split into two fragments. Lines connecting the qubits (illustrated as atoms) indicate general spin interactions $\hat{S}^{(i)}_{\alpha} \hat{S}^{(j)}_{\beta}$ acting on qubits $i$ and $j$, weighted by $J_{ij}^{\alpha, \beta}$. The interactions highlighted in red are omitted due to fragmentation; these form the ``interface'' of the two fragments. An auxiliary qubit $a$ interacts with the qubits within a fragment, according to the prescribed interactions of the circled target qubit in the opposite fragment. While, for visual clarity, this diagram only displays the mean-field corrections applied to auxiliary $a_{1}$, in reality, such corrections are applied to each qubit that participates in one or more interactions beyond the fragment boundary.}
    \label{fig:fragment-cartoon}
\end{figure*}

For added clarification, here we consider a toy model of $N = 6$ qubits fragmented into two groups such that $N_f = 3$ (see Fig. \ref{fig:fragment-cartoon}, a more explicitly labeled version of Fig. 1b). One auxiliary qubit is included for each fragment ($N_a = 1$). The auxiliary qubit of Fragment 1, $a_1$, targets qubit 5 in Fragment 2. Similarly, the auxiliary qubit of Fragment 2, $a_2$, targets qubit 1 in Fragment 1. In the figure, the interactions mediated by the auxiliary qubits are explicitly labeled to illustrate how each auxiliary qubit plays the role of one qubit from the environment. Which environment qubit is selected for this role will depend on the quantum variance $\text{var}( H - H_{I}^{(f)})$ (see Section \ref{sec:aux-selection}). The mean-field corrections applied to $a_1$ are included in the figure to account for the missing interactions with qubits 4 and 6 that the target qubit (qubit 5) would participate in if all qubits were present. Although only one such arrow appears in the figure, these corrections exist for each qubit participating in fewer interactions than it would in the full system, regardless of whether the qubit is categorized as a fragment qubit or an auxiliary targeting an environment qubit. 

\subsection{Fragmentation Error}\label{sec:frag-error}
Consider the error caused by the omission of interactions that form the interface of some particular fragment $f$ and the remaining environment spins $E$. This Hamiltonian which we denote $H_{I}^{(f)}$ is subtly different than the previously introduced $H^{(f)}$, as it includes operators acting in the space of $E$ in order to isolate the error caused by the section of the interaction interface produced by a particular fragment $f$: 
\begin{equation}
\begin{split}
    H_{I}^{(f)} &= H^{(f)} + H^{(E)} \\
    &= -\sum_{\langle i,j \rangle /\in I}
    \sum_{\alpha, \beta} J_{ij}^{\alpha, \beta} \hat{S}^{(i)}_{\alpha} \hat{S}^{(j)}_{\beta} -  \sum_{i}^{N} h_{i} \hat{S}^{(i)}_x .
\end{split}
\end{equation}
The fidelity of the state evolved under $H_{I}^{(f)}$ with that evolved under $H$ is given by: 
\begin{equation}
    F(t) = |\langle \Psi| U^{\dagger}(t) U_{I}^{(f)}(t) |\Psi \rangle|^2 ,
\end{equation}
where $U(t) = \exp(-iHt)$, $U_{I}^{(f)}(t) = \exp(-iH_{I}^{(f)}t)$, and $|\Psi_0\rangle$ is the current state of the system \cite{liang_quantum_2019}.

To analyze the fidelity metric, we can combine the product $U^{\dagger}(t) U_{I}^{(f)}(t)$ into a single exponential $\exp({Z})$. This final exponential argument is obtained using the Baker–Campbell–Hausdorff (BCH) formula \cite{gilmore_bakercampbellhausdorff_1974}, which provides an expression for $Z$ in terms of the nested commutators of the individual exponential arguments of $U^{\dagger}(t)$ and $U_{I}^{(f)}(t)$: 
\begin{equation}
    \label{eq:bch}
    \begin{split} 
    Z &= it(H - H_{I}^{(f)}) + \frac{1}{2}t^2[H, H_{I}^{(f)}] + \frac{1}{12}t^3 \big( i[H, [H, H_{I}^{(f)}]] + (-i)[H_{I}^{(f)}, [H_{I}^{(f)}, H]] \big)
    \\[.5ex] &\qquad - \frac{1}{24}t^4 [H_{I}^{(f)}, [H, [H, H_{I}^{(f)}]]] + \cdots
    \end{split}
\end{equation}
Notice that each subsequent term in the BCH formula has a higher order in $t$. To approximate the error, we keep the terms up to small orders in $t$ -- specifically, we can write $Z = \sum_{n=1}^{\infty} z_{n} t^{n}$, where the coefficients $z_{n}$ do not depend on $t$ and generally do not commute with each other, and truncate the sum after some $n$. We note that the first and second order $z_{n}$ are given by: 
\begin{equation}
    z_1 = i(H-H_{I}^{(f)}) ,
\end{equation}
\begin{equation}
    z_2 = \frac{1}{2}[H, H_{I}^{(f)}] .
\end{equation}
Taylor expanding the exponential $\exp({Z})$ to second order in $t$:
\begin{equation}
    \begin{split} 
    U^{\dagger}(t) U_{I}^{(f)}(t) &= \exp\Big( \sum_{n=1}^{\infty} z_{n} t^{n} \Big)
    \\[.5ex] 
    &= 1 + \sum_{n=1}^{\infty} z_{n} t^{n} + \frac{1}{2}\Big( \sum_{n=1}^{\infty} z_{n} t^{n} \Big) ^2  + \cdots 
    \\[.5ex]
    &= 1 + z_{1}t + t^2 \Big(z_{2} + \frac{1}{2} z_{1}^2\Big) + \mathcal{O}(t^3) .
    \end{split}
\end{equation}

The fidelity is then the modulus square of the expectation of the above expression, taken with respect to the initial state of the full system: 
\begin{equation}
    \begin{split}
    F(t) &\approx \big|1 + \langle z_{1} \rangle t + t^2 \big( \langle z_{2} \rangle + \frac{1}{2}\langle z_1 ^2 \rangle  \big) \big|^2    
    \\[.5ex] &= 1 + t \big( \langle z_{1} \rangle + \langle z_{1} \rangle ^*\big) + t^2 \big( |\langle z_{1} \rangle|^2 + \langle z_{2} \rangle + \langle z_{2} \rangle^* + \frac{1}{2}\langle z_{1}^2 \rangle + \frac{1}{2}\langle z_{1}^2 \rangle^*\big) + \mathcal{O}(t^{3}) .
    \end{split}
\end{equation}
Plugging in the expressions for $z_1$ and $z_2$: 
\begin{equation}
    \begin{split}
    F(t) &\approx 1 + t\big(i \langle H - H_{I}^{(f)} \rangle -i \langle H - H_{I}^{(f)} \rangle ^* \big) + t^2 \big( |\langle H - H_{I}^{(f)} \rangle|^2  + \frac{1}{2} \langle [H, H_{I}^{(f)}] \rangle + \frac{1}{2} \langle [H, H_{I}^{(f)}] \rangle ^* 
    \\[.5ex] &\qquad + (i)^2 \frac{1}{2}\langle (H - H_{I}^{(f)})^2 \rangle + (-i)^2 \frac{1}{2} \langle (H - H_{I}^{(f)})^2 \rangle ^*\big) .
    \end{split}
\end{equation}
Notice that, as both $H$ and $H_{I}^{(f)}$ are Hermitian, $\langle H - H_{I}^{(f)} \rangle ^* = \langle H \rangle ^* - \langle H_{I}^{(f)} \rangle ^* = \langle H \rangle - \langle H_{I}^{(f)} \rangle$. Thus, the terms first order in $t$ cancel with one another. The second-order terms can be simplified by expanding them: 
\begin{equation}
    \begin{split}
    F(t) &\approx 1 + t^2 \big( |\langle H - H_{I}^{(f)} \rangle|^2 + \frac{1}{2} (\langle H H_{I}^{(f)}\rangle - \langle H_{I}^{(f)} H\rangle) + \frac{1}{2} (\langle H H_{I}^{(f)}\rangle^* - \langle H_{I}^{(f)} H\rangle^*) 
    \\[.5ex] &\qquad - \frac{1}{2} (\langle H^2 \rangle - \langle H H_{I}^{(f)} \rangle - \langle H_{I_f} H\rangle + \langle (H_{I}^{(f)})^{2} \rangle) 
    \\[.5ex] &\qquad - \frac{1}{2} (\langle H^2 \rangle^* - \langle H H_{I}^{(f)} \rangle^* - \langle H_{I}^{(f)} H \rangle^* + \langle (H_{I}^{(f)})^{2} \rangle^*)\big) .
    \end{split}
\end{equation}
Using the fact that $\langle A B \rangle^* = \langle B^{\dagger} A^{\dagger} \rangle$ to simplify, we arrive at a compact expression for the fidelity to second order in $t$: 
\begin{equation}
    \begin{split}
    F(t) &= 1 + t^2 \big( |\langle H - H_{I}^{(f)} \rangle|^2 - \langle (H - H_{I}^{(f)})^2 \rangle \big) + \mathcal{O}(t^3)
    \\[.5ex]&\approx 1 - \text{var}( H - H_{I}^{(f)}) t^2 .
    \end{split}
\end{equation}
The error $\epsilon(t) = 1 - F(t)$ is thus given by $\text{var}( H - H_{I}^{(f)}) t^2$ for short times $t$. 

\subsection{Numerical Simulation of a Quantum Channel}\label{sec:Jij}
When simulators are linked via a quantum channel, the states of the fragments no longer live in separate Hilbert spaces; they comprise the state of the collective system, which lives in the larger Hilbert space of size $2^N$. To simulate this numerically, we evolve a modified Hamiltonian that lives in the full Hilbert space. This Hamiltonian involves adjusted connectivity matrices $J_{ij}^{\alpha\beta}$ to match the connectivity provided by the current auxiliary encoding. That is, defining a set $S_f$ for each fragment containing the list of fragment qubits and target qubits for auxiliaries of fragment $f$:
\begin{equation}
    S_f = \{i \in f\} \cup \{a_f\} ,
\end{equation}
we implement the union of connectivity matrices $\cup_f J_{ij}^{\alpha\beta, (f)}$ for each $\alpha\beta$ interaction type, defined by: 
\begin{equation}
    J_{ij}^{\alpha\beta, (f)} = \begin{cases}
      J_{ij}^{\alpha\beta} & \text{if $i, j \in S_f$}\\
      0 & \text{otherwise} .
    \end{cases}       
\end{equation}
Any interactions still zeroed in the union $\cup_f J_{ij}^{\alpha\beta, (f)}$ are approximately included via mean-field corrections. If the auxiliary encoding is updated, the union $\cup_f J_{ij}^{\alpha\beta, (f)}$ will change due to the change in selected auxiliaries $\{a_f\}$ of each fragment.

\section{When Mean-Field Corrections Are Beneficial to Time Evolution}\label{sec:benefit-of-mf}
In this appendix, we numerically investigate the role of mean-field corrections in time evolution, showing that for the majority of distributed states, mean-field corrections reduce the fragmentation error.
The second order fragmentation error is derived in the main text to be the variance of the difference between $H$ and $H_{I}^{(f)}$ (see Eq. (\ref{eq:var})). In this appendix, we will denote this quantity by $V$. If we include mean-field corrections in the fragmented Hamiltonian, the second order fragmentation error can be amended to include additional terms: 
\begin{equation} \label{eq:var-mf}
\begin{split}
    V_{MF}&\coloneqq \text{var}(H - H_{I, MF}^{(f)}) 
    \\&= var\bigg( -\sum_{\langle i,j \rangle \in I} \sum_{\alpha, \beta} J_{ij}^{\alpha, \beta} \hat{S}^{(i)}_{\alpha} \big(\hat{S}^{(j)}_{\beta} - \langle \hat{S}^{(j)}_{\beta}\rangle \big) \bigg) .
\end{split} 
\end{equation}
The improved performance due to mean-field corrections present in the results of the main text can be attributed to the fact that in those scenarios, $V_{MF} < V$ on average, such that the error is decreased. However, situations can arise where $V_{MF} > V$. In what follows, we show that these situations generally occur when the qubits spanning the interface are significantly entangled with one another. 

\begin{figure*}[ht]
    \centering
    \includegraphics[width=3.5in]{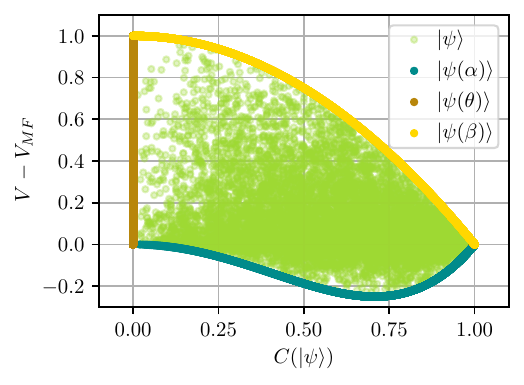}
    \caption{Variance difference $V - V_{MF}$ versus concurrence $C(|\psi\rangle)$ for 10,000 randomly generated two-qubit states $|\psi\rangle$. These quantities for specific parameterized states $|\psi(\alpha)\rangle$, $|\psi(\theta)\rangle$, and $|\psi(\beta)\rangle$ (defined in Eqs. (\ref{eq:alpha}) -- (\ref{eq:beta})) are plotted in separate colors. To calculate the variance difference, a single interaction $\hat{\sigma}_z^{(1)}\hat{\sigma}_z^{(2)}$ is considered.}
    \label{fig:scatter-v-diff}
\end{figure*}

To gain insight into when mean-field corrections are beneficial, consider two fragments each containing a single qubit. The qubits interact according to a quantum Ising-like Hamiltonian, with $J_{12} = 4$ such that $J_{12}\hat{S}_{z}^{(1)}\hat{S}_{z}^{(2)}$ reduces to Pauli operators $\hat{\sigma}_{z}^{(1)}\hat{\sigma}_{z}^{(2)}$. The interface of the fragments is comprised of the single $J_{12}$ linking the two qubits of the system. To explore the possible values of $V - V_{MF}$, we randomly generate 10,000 two-qubit states $|\psi\rangle$:
\begin{equation}
    |\psi\rangle = c_{00}|00\rangle + c_{01}|01\rangle + c_{10}|10\rangle + c_{11}|11\rangle,
\end{equation}
where the complex coefficients $c_{ij}$ are properly normalized. For each $|\psi\rangle$, the variance difference $V - V_{MF}$ is computed (recall that a positive $V - V_{MF}$ implies that mean-field corrections have reduced the simulation error). We additionally compute the concurrence to measure the entanglement between qubit 1 and qubit 2. The concurrence is defined to be \cite{miranowicz_ordering_2004}: 
\begin{equation}
    C(|\psi\rangle) = 2|c_{00}c_{11} - c_{01}c_{10}| ,
\end{equation}
where $C(|\psi\rangle) = 0$ for any pure state $|\psi\rangle$, and $C(|\psi\rangle)$ monotonically increases with entanglement to a maximum value of $C(|\psi\rangle) = 1$ (e.g., for a maximally entangled Bell pair). The results are plotted in light green in Fig. \ref{fig:scatter-v-diff}. The scattered data falls within a closed area, with $V - V_{MF}$ dropping below zero for concurrence $0 < C(|\psi\rangle) < 1$. The difference $V - V_{MF}$ is also bounded between $-0.25$ and $1$, reaching its maximum positive value for product states (where $C(|\psi\rangle)$).

To understand the features of a state that extremizes $V - V_{MF}$, we probe the boundaries of the envelope of possible states by examining parameterized two-qubit states. The specific parameterized states discussed in the following paragraphs are meant to be representative of the form of the states that lie on the edge of the closed area in Fig. \ref{fig:scatter-v-diff}, but generally, many possible two-qubit states will produce identical values of $(C(|\psi\rangle), V-V_{MF})$. 

\begin{figure*}[ht]
    \centering
    \includegraphics[width=6.0in]{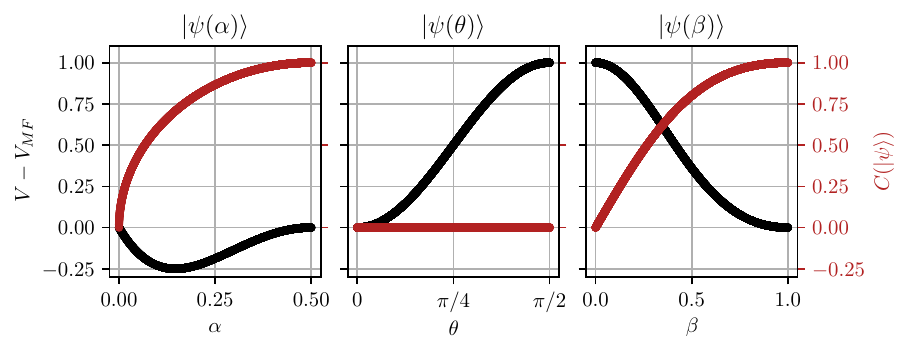}
    \caption{Variance difference $V - V_{MF}$ (left axis) and concurrence $C(|\psi\rangle)$ (right axis) plotted versus a state parameter for specific parameterized two-qubit states, specified by the subplot title. The parameterized states are defined in Eqs. (\ref{eq:alpha}) -- (\ref{eq:beta}).}
    \label{fig:parameterized-v-diff}
\end{figure*}

First, we consider a parameterized state $|\psi(\alpha)\rangle$ that smoothly approaches the Bell state $|\Phi_{+}\rangle = 1/\sqrt{2}(|00\rangle + |11\rangle)$:
\begin{equation}\label{eq:alpha}
    |\psi(\alpha)\rangle \coloneqq \sqrt{\alpha} |00\rangle + \sqrt{1 - \alpha}|11\rangle .
\end{equation}
When $\alpha = 0$, the resulting state $|11\rangle$ is a product state with zero concurrence. Furthermore, $|11\rangle$ is computational basis state with $V = V_{MF} = 0$. As $\alpha$ approaches $1/2$, the concurrence increases, reaching a maximum value. The variance difference, on the other hand, decreases to take on negative values -- see the left panel of Fig. \ref{fig:parameterized-v-diff}. Thus, it is states with structured entanglement such as $|\psi(\alpha)\rangle$ which will not benefit from mean-field corrections. Plotting $V - V_{MF}$ versus the concurrence of $|\psi(\alpha)\rangle$ in Fig. \ref{fig:scatter-v-diff}, we see that this set of states lies on the lower boundary of the envelope of possible values.

Secondly, we consider the two-qubit state produced by rotating qubit 1 of the state $|00\rangle$ about $x$ by an angle $\theta$:
\begin{equation}\label{eq:theta}
    |\psi(\theta)\rangle \coloneqq \cos{(\theta/2)} |00\rangle - i \sin{(\theta/2)}|10\rangle .
\end{equation}
The resulting state will always be a product state with zero concurrence; however, the variance difference will depend on $\theta$. Sweeping across $\theta$ (see the middle panel of Fig. \ref{fig:parameterized-v-diff}), we see that $V-V_{MF}$ grows as $\theta$ is increased. A state of this kind lies on the left boundary of the envelope (see Fig. \ref{fig:scatter-v-diff}). 

Finally, we consider a state that lies on the top boundary of the envelope, with positive values for both the variance difference and the concurrence. We parameterize this by interpolating between a product state with maximum $V - V_{MF}$ and a state with maximum concurrence while avoiding a Bell state, which is known to lie on the lower boundary:
\begin{equation}\label{eq:beta}
    |\psi(\beta)\rangle \coloneqq \frac{1}{\sqrt{2 + 2\beta^2}}\big( |00\rangle + \beta|01\rangle - |10\rangle + \beta|11\rangle \big) .
\end{equation}
The right panel of Fig. \ref{fig:parameterized-v-diff} reveals that $V-V_{MF}$ and $C(|\psi(\beta)\rangle)$ remain non-negative as $\beta$ is varied. Again, plotting $V - V_{MF}$ versus the concurrence of $|\psi(\beta)\rangle$ in Fig. \ref{fig:scatter-v-diff} reveals that such a state indeed lies at the top boundary of the envelope. 

In the time evolution applications considered in the main text, all initial states are computational basis states. It is thus unlikely that the state evolves into one with large entanglement across the interface with the particular structure required to produce a negative value of $V - V_{MF}$. This is why, on average, including mean-field corrections enhances the performance of the scheme. More generally, we expect that in practical situations where fragmentation is employed, it is unlikely that highly structured entanglement will form across a fragment interface through evolution. Therefore, it is reasonable to expect that mean-field corrections will still improve performance in general applications. 

\section{Adding a Perturbative Transverse Field to the Ising Model}\label{sec:perturbation-theory}

Consider perturbing an Ising-like Hamiltonian with a small transverse field $h$. We will first derive how the perturbation shifts the ground state energy and the corresponding eigenstate for the case of an unperturbed Hamiltonian containing all Ising-like interactions. Then, we will derive the shifts for the unconventional situation where certain interactions within the Hamiltonian are replaced by mean-field corrective terms. Understanding how these cases differ from each other will illuminate how the ground state solution of a mean-field corrected Hamiltonian strays from that of the full Hamiltonian as $h$ grows, rendering the Ising-like model increasingly quantum. 

We write down the Hamiltonian containing all interactions as: 
\begin{equation}
    H = H_{0} + H_{I} + \lambda V, 
\end{equation}
where $H_0$ contains the intra-fragment interactions:
\begin{equation}
    H_{0} = -\sum_{\langle i, j \rangle \notin I} J_{ij} \hat{S}_{z}^{(i)} \hat{S}_{z}^{(j)} ,
\end{equation}
$H_I$ contains the inter-fragment interactions:
\begin{equation}
    H_{I} = -\sum_{\langle i, j \rangle \in I} J_{ij} \hat{S}_{z}^{(i)} \hat{S}_{z}^{(j)} , 
\end{equation}
$V$ contains the perturbing transverse field:
\begin{equation}
    V = -h \sum_{i} \hat{S}_{x}^{(i)} ,
\end{equation}
and $\lambda$ is a perturbation parameter. We explicitly separate the interactions that will be replaced by mean-field corrections; these terms comprise $H_{I}$. 

In contrast, the mean-field corrected Hamiltonian denoted $H_{MF}$ is given by: 
\begin{equation}
    H_{MF}(|\psi\rangle) = H_{0} + H_{I, MF}(|\psi\rangle) + \lambda V , 
\end{equation}
where the form of the Hamiltonian now depends on the state of the system due to the mean-field corrections:
\begin{equation}
    H_{I, MF}(|\psi\rangle) = -\sum_{\langle i, j \rangle \in I} J_{ij} \sigma_{z}^{(i)} \langle \psi | \sigma_{z}^{(j)} | \psi \rangle .
\end{equation}
This state-dependence must be taken into account when applying perturbation theory.
% [Still trying to determine if it should look like this instead:]
% \begin{equation}
%     H_{I, MF}(|\psi\rangle) = -\sum_{\langle i, j \rangle \in I} \frac{J_{ij}}{2} \bigg( \sigma_{z}^{(i)} \langle \psi | \sigma_{z}^{(j)} | \psi \rangle + \langle \psi | \sigma_{z}^{(i)} | \psi \rangle \sigma_{z}^{(j)} \bigg) .
% \end{equation}

\subsection{Perturbing the Full Hamiltonian}
In this section, we follow the degenerate perturbation theory procedure outlined in \cite{sakurai_modern_2020}. When all interactions are included, the unperturbed Hamiltonian is given by the sum of $H_0$ and $H_I$. The eigendecomposition of this operator defines the zeroth order eigenenergies and eigenstates:
\begin{equation}
    H_{0} +  H_{I} = \sum_{k} E_{k}^{(0)} |k^{(0)}\rangle \langle k^{(0)} | .
\end{equation}
The unperturbed Hamiltonian is classical, with computational basis states for eigenstates. It also possesses $\mathbb{Z}_2$ symmetry, such that each eigenstate $|x\rangle$ and its ``flipped'' version $|\bar{x}\rangle$ are degenerate, including the ground state $|x^{*}\rangle$. It is necessary to find the correct linear combinations of the degenerate states to properly calculate the higher order corrections. Typically, this is accomplished by diagonalizing the perturbing Hamiltonian $V$ in the subspace of degenerate eigenstates, which yields the proper zeroth order eigenstates as well as the first order energy corrections, $\{E_{k}^{(1)}\}$. The particular $V$ of this problem -- a global transverse field -- vanishes in the subspace of $|x\rangle$, $|\bar{x}\rangle$, requiring the first order energy corrections $\{E_{k}^{(1)}\}$ to vanish but providing no insight into the correct zeroth order eigenstates. 

To calculate the first order eigenstate corrections, the correction is split into one within the degenerate space $D_{k}$ and one in the space outside $D_{k}$, which we define as $\bar{D}_k$. For the latter correction, we have:
\begin{equation}
    P_{\bar{D}_{k}} | k^{(1)} \rangle = \sum_{m \notin D_{k}} \frac{\langle m^{(0)} |V| k^{(0)} \rangle}{E_{k}^{(0)} - E_{m}^{(0)}} |m^{(0)} \rangle ,
\end{equation}
where $P_{\bar{D}_{k}}$ is the projects out the degenerate subspace. Following \cite{sakurai_modern_2020}, the correction within the degenerate subspace is given by: 
\begin{equation}
    P_{D_{k}} | k^{(1)} \rangle = \sum_{\substack{k' \in D_{k}, \\ k' \neq k}} \frac{P_{D_{k}}|k'^{(0)}\rangle}{E_{k}^{(1)} - E_{k'}^{(1)}} \langle k'^{(0)} | V P_{\bar{D}_k} \frac{1}{E_{k}^{(0)} - H_0 - H_I}  P_{\bar{D}_k} V |k^{(0)} \rangle .
\end{equation}
The first order energy difference in the denominator of the first fraction of this expression is singular, because the degeneracy within $D_{k}$ has not been lifted by the first order energy correction. However, this apparent issue provides us with the correct zeroth order eigenstates required to cancel the singular denominator: $\{|k^{(0)}\rangle\}$ must be selected to diagonalize the object $W := V P_{\bar{D}_k} (E_{k}^{(0)} - H_0 - H_I)^{-1}  P_{\bar{D}_k} V$ within the subspace of degenerate states, such that the quantity $\langle k'^{(0)} | W |k^{(0)} \rangle$ vanishes for all $k' \neq k$ within $D_k$. With this choice for $\{|k^{(0)}\rangle\}$, the first order eigenstate correction within the degenerate space vanishes, and the full first order eigenstate correction takes the form: 
\begin{equation}
    | k^{(1)} \rangle = \sum_{m \notin D_{k}} \frac{\langle m^{(0)} |V| k^{(0)} \rangle}{E_{k}^{(0)} - E_{m}^{(0)}} |m^{(0)} \rangle .
\end{equation}

One can show that $\langle x | W | x \rangle = \langle x | W | \bar{x} \rangle = \langle \bar{x} | W | x \rangle = \langle \bar{x} | W | \bar{x} \rangle$ using the $\mathbb{Z}_2$ symmetry of the unperturbed Hamiltonian. Thus, the correct zeroth order eigenstates are the symmetric and antisymmetric superpositions of $|x\rangle$ and $|\bar{x}\rangle$, given by $|\pm_{x}\rangle = (1/\sqrt{2})(|x\rangle \pm |\bar{x}\rangle)$.

% \begin{equation}
%     E_{l}^{(2)} = \sum_{k \notin D_{k}} \frac{|\langle k^{(0)} |V| l^{(0)} \rangle|^{2}}{E_{l}^{(0)} - E_{k}^{(0)}}
% \end{equation}

\subsection{Perturbing the Mean-Field Corrected Hamiltonian}
Throughout this derivation, we will take advantage of the particular form of $H_{I, MF}(|\psi\rangle)$ to make simplifications. 

Consider a modified Schrödinger equation that takes into account the state-dependence of $H_{MF}(|\psi\rangle)$: 
\begin{equation}
    (H_{0} + H_{I, MF}(|k_{MF}\rangle) + \lambda V) |k_{MF}\rangle = E_{k,MF}|k_{MF}\rangle
\end{equation}
In conventional perturbation theory, the eigenstates and eigenvalues ($|k_{MF}\rangle$ and $E_{k, MF}$, respectively) are expanded about their unperturbed counterparts, $|k_{MF}^{(0)}\rangle$ and $E_{k, MF}^{(0)}$. However, the modified Schrödinger equation written above is no longer an eigenvalue problem; the dependence on $|k_{MF}\rangle$ is non-linear. In the spirit of conventional perturbation theory, we will proceed in the usual manner, taking special care to include the non-linearity. Expanding $|k_{MF}\rangle$ and $E_{k, MF}$ in orders of $\lambda$: 
\begin{equation}
    |k_{MF}\rangle = |k_{MF}^{(0)}\rangle + \lambda|k_{MF}^{(1)}\rangle + \lambda^2 |k_{MF}^{(2)}\rangle + \cdots ,
\end{equation}
\begin{equation}
    E_{k, MF} = E_{k, MF}^{(0)} + \lambda E_{k, MF}^{(1)} + \lambda^2 E_{k, MF}^{(2)} + \cdots .
\end{equation}
Replacing $|k_{MF}\rangle$ and $E_{k, MF}$ in the modified Schrödinger equation: 
\begin{align*}
    \Big( H_{0} &+ H_{I, MF}\big(|k_{MF}^{(0)}\rangle + \lambda|k_{MF}^{(1)}\rangle + \lambda^2 |k_{MF}^{(2)}\rangle + \cdots\big) + \lambda V \Big) \big( 
    |k_{MF}^{(0)}\rangle + \lambda|k_{MF}^{(1)}\rangle + \lambda^2 |k_{MF}^{(2)}\rangle + \cdots \big) 
    \\ &= \big( E_{k, MF}^{(0)} + \lambda E_{k, MF}^{(1)} + \lambda^2 E_{k, MF}^{(2)} + \cdots\big)  \big( |k_{MF}^{(0)}\rangle + \lambda|k_{MF}^{(1)}\rangle + \lambda^2 |k_{MF}^{(2)}\rangle + \cdots \big)
\end{align*}
Now, we can isolate and equate the various orders of $\lambda$ to calculate the perturbations. If we first consider the zeroth order, we recover the unperturbed spectrum: 
\begin{equation}
    \big(H_{0} + H_{I, MF}(|k_{MF}^{(0)}\rangle)\big) |k_{MF}^{(0)}\rangle = E_{k, MF}^{(0)} |k_{MF}^{(0)}\rangle
\end{equation}
Moving to higher orders, it is helpful to write the explicit form of $H_{I, MF}(|\psi\rangle)$ to properly count the orders of $\lambda$. The first order terms take the form: 
\begin{equation}\label{eq:1st-order-MF-PT}
 \begin{aligned}
    H_{0} |k_{MF}^{(1)}\rangle &+ V |k_{MF}^{(0)}\rangle 
    -\sum_{\langle i, j \rangle \in I} J_{ij} \bigg( \sigma_{z}^{(i)} \langle k_{MF}^{(0)} | \sigma_{z}^{(j)} | k_{MF}^{(0)} \rangle \bigg) |k_{MF}^{(1)}\rangle 
    \\ & -\sum_{\langle i, j \rangle \in I} J_{ij} \bigg( \sigma_{z}^{(i)} \langle k_{MF}^{(1)} | \sigma_{z}^{(j)} | k_{MF}^{(0)} \rangle + \sigma_{z}^{(i)} \langle k_{MF}^{(0)} | \sigma_{z}^{(j)} | k_{MF}^{(1)} \rangle \bigg) |k_{MF}^{(0)}\rangle 
    \\ &= E_{k, MF}^{(0)} |k_{MF}^{(1)}\rangle + E_{k, MF}^{(1)} |k_{MF}^{(0)}\rangle .
\end{aligned}   
\end{equation}
The operators $\sigma_{z}^{(j)}$ are diagonal in the computational basis $\{|k_{MF}^{(0)}\rangle\}$. By definition, corrections to $|k_{MF}^{(0)}\rangle$ such as $|k_{MF}^{(1)}\rangle$ will be orthogonal to $|k_{MF}^{(0)}\rangle$; therefore, terms of the form $\langle k_{MF}^{(1)} | \sigma_{z}^{(j)} | k_{MF}^{(0)} \rangle $ will vanish. The first order equation thus simplifies to: 
\begin{equation}
    H_{0} |k_{MF}^{(1)}\rangle + V |k_{MF}^{(0)}\rangle 
    -\sum_{\langle i, j \rangle \in I} J_{ij} \bigg( \sigma_{z}^{(i)} \langle k_{MF}^{(0)} | \sigma_{z}^{(j)} | k_{MF}^{(0)} \rangle \bigg) |k_{MF}^{(1)}\rangle = E_{k, MF}^{(0)} |k_{MF}^{(1)}\rangle + E_{k, MF}^{(1)} |k_{MF}^{(0)}\rangle .
\end{equation}
To determine the first order energy shift, we project the above equation with the unperturbed eigenstate $\langle k_{MF}^{(0)}|$:
\begin{equation}\label{eq:H0-off-diagonal}
\begin{aligned}
    \langle k_{MF}^{(0)}| H_{0} |k_{MF}^{(1)}\rangle &+ \langle k_{MF}^{(0)}|V |k_{MF}^{(0)}\rangle 
    -\sum_{\langle i, j \rangle \in I} J_{ij} \langle k_{MF}^{(0)}|\sigma_{z}^{(i)} |k_{MF}^{(1)}\rangle \langle k_{MF}^{(0)} | \sigma_{z}^{(j)} | k_{MF}^{(0)} \rangle 
    \\&= E_{k, MF}^{(0)} \langle k_{MF}^{(0)}|k_{MF}^{(1)}\rangle + E_{k, MF}^{(1)} \langle k_{MF}^{(0)}|k_{MF}^{(0)}\rangle .
\end{aligned}    
\end{equation}
The terms originating from $H_{I, MF}(|\psi\rangle)$ as well as the first term on the right-hand side of the equation will vanish due to the previously discussed fact that $|k_{MF}^{(1)}\rangle$ will be orthogonal to $|k_{MF}^{(0)}\rangle$. Let us examine the first term on the left-hand side, making use of the definition of the unperturbed spectrum: 
\begin{align*}
    \langle k_{MF}^{(0)}| H_{0} |k_{MF}^{(1)}\rangle &= \langle k_{MF}^{(0)}| E_{k, MF}^{(0)} - H_{I, MF}(|k_{MF}^{0}\rangle) |k_{MF}^{(1)}\rangle
    \\ &= E_{k, MF}^{(0)} \langle k_{MF}^{(0)} | k_{MF}^{(1)}\rangle + \sum_{\langle i, j \rangle \in I} J_{ij} \langle k_{MF}^{(0)}|\sigma_{z}^{(i)} |k_{MF}^{(1)}\rangle \langle k_{MF}^{(0)} | \sigma_{z}^{(j)} | k_{MF}^{(0)} \rangle 
    \\ &= 0 .
\end{align*}
Thus, this term vanishes for the same reason, and the first order energy shift from conventional perturbation theory is recovered: 
\begin{equation}
    E_{k, MF}^{(1)}  = \langle k_{MF}^{(0)}|V |k_{MF}^{(0)}\rangle .
\end{equation}
Of course, this energy correction vanishes for all $k$, in the same way that the first order energy shift of the full Hamiltonian vanishes. 

%Before broaching the derivation of $E_{k}^{(2)}$, let us derive the first order correction to the eigenstate, $|k^{(1)}\rangle$. 
To determine the first order eigenstate shift $|k_{MF}^{(1)}\rangle$, we project the first order equation given in Eq. \ref{eq:1st-order-MF-PT} with the unperturbed eigenstate $\langle m_{MF}^{(0)}|$, where $m \neq k$:
\begin{equation}\label{eq:1st-order-MF-PT-2}
\begin{aligned}
    \langle m_{MF}^{(0)}| H_{0} |k_{MF}^{(1)}\rangle &+ \langle m_{MF}^{(0)}|V |k_{MF}^{(0)}\rangle 
    -\sum_{\langle i, j \rangle \in I} J_{ij} \langle m_{MF}^{(0)}|\sigma_{z}^{(i)} |k_{MF}^{(1)}\rangle \langle k_{MF}^{(0)} | \sigma_{z}^{(j)} | k_{MF}^{(0)} \rangle 
    \\&= E_{k, MF}^{(0)} \langle m_{MF}^{(0)}|k_{MF}^{(1)}\rangle + E_{k, MF}^{(1)} \langle m_{MF}^{(0)}|k_{MF}^{(0)}\rangle .
\end{aligned}  
\end{equation}
The only term guaranteed to vanish is the second term on the right-hand side, as $\langle m_{MF}^{(0)}|k_{MF}^{(0)}\rangle = 0$ for $m \neq k$. The term $\langle m_{MF}^{(0)}| H_{0} |k_{MF}^{(1)}\rangle$ can also be simplified, with careful attention to the definition of the zeroth order energy $E_{m, MF}^{(0)}$: 
\begin{equation}
\begin{aligned}
    \langle m_{MF}^{(0)}| H_{0} |k_{MF}^{(1)}\rangle &= \langle m_{MF}^{(0)}| E_{m, MF}^{(0)} - H_{I, MF}(|k_{MF}^{0}\rangle) |k_{MF}^{(1)}\rangle
    \\ &= E_{m, MF}^{(0)} \langle m_{MF}^{(0)} | k_{MF}^{(1)}\rangle + \sum_{\langle i, j \rangle \in I} J_{ij} \langle m_{MF}^{(0)}|\sigma_{z}^{(i)} |k_{MF}^{(1)}\rangle \langle k_{MF}^{(0)} | \sigma_{z}^{(j)} | k_{MF}^{(0)} \rangle .
\end{aligned}
\end{equation}
Notice that the mean-field terms in the expression above precisely cancel with the mean-field terms remaining in Eq. \ref{eq:1st-order-MF-PT-2}. Simplifying: 
\begin{equation}
    \langle m_{MF}^{(0)}|V |k_{MF}^{(0)}\rangle = (E_{k, MF}^{(0)} - E_{m, MF}^{(0)}) \langle m^{(0)}|k^{(1)}\rangle .
\end{equation}
Therefore, the first order correction to the eigenstate is identical to that of the full Hamiltonian with no mean-field corrections: 
\begin{equation}
\begin{aligned}
    |k_{MF}^{(1)}\rangle &= \sum_{m} |m_{MF}^{(0)}\rangle \langle m_{MF}^{(0)} |k_{MF}^{(1)}\rangle 
    \\ &= \sum_{m \notin D_{k}} |m_{MF}^{(0)}\rangle \frac{\langle m_{MF}^{(0)}|V |k_{MF}^{(0)}\rangle}{E_{k, MF}^{(0)} - E_{m, MF}^{(0)}} .
\end{aligned}
\end{equation}

\subsection{First Order Eigenstate Overlap}\label{sec:overlap}
In this section, we derive the overlap between an eigenstate of an Ising-like Hamiltonian $H$ with a small transverse field and the corresponding eigenstate of a mean-field corrected version of the Hamiltonian, $H_{MF}$, to leading order in perturbation theory. This overlap is approximately correct as long as the strength of the transverse field $|h|$ is small enough that first order perturbation theory is a good description of the eigenstates.

To first order, an eigenstate of each Hamiltonian is given by:
\begin{equation}
    |\psi_{k}\rangle = \mathcal{N}\big( |k^{(0)} \rangle + |k^{(1)} \rangle \big) ,
\end{equation}
\begin{equation}
    |\psi_{k, MF}\rangle = \mathcal{N}_{MF}\big( |k_{MF}^{(0)} \rangle + |k_{MF}^{(1)} \rangle \big) ,
\end{equation}
where $\mathcal{N}$, $\mathcal{N}_{MF}$ are normalization factors. The zeroth order eigenstates are normalized by definition, so re-normalization is required when any higher order corrections are included. 

The overlap between the two states is given by their inner product:
\begin{equation}
    \langle \psi_{k} | \psi_{k, MF} \rangle = \mathcal{N}^{*}\mathcal{N}_{MF} \big( \langle k^{(0)} | k_{MF}^{(0)} \rangle + \langle k^{(1)} | k_{MF}^{(1)} \rangle \big) ,
\end{equation}
where the cross-terms $\langle k^{(1)} | k_{MF}^{(0)} \rangle$, $\langle k^{(0)} | k_{MF}^{(1)} \rangle$ have been dropped due to the fact that the first order corrections $| k^{(1)} \rangle$, $| k_{MF}^{(1)} \rangle$ lie outside the degenerate subspace of their corresponding zeroth order states $| k^{(0)} \rangle$ and $| k_{MF}^{(0)} \rangle$, which both live in the same degenerate subspace $D_k$.

As argued in previous sections, the zeroth order eigenstates $|k_{MF}^{(0)}\rangle$ are computational basis states $|x\rangle$, while the zeroth order eigenstates $|k^{(0)}\rangle$ are superpositions of two computational basis states related by $\mathbb{Z}_2$ symmetry, $|\pm_{x}\rangle = (1/\sqrt{2})(|x\rangle \pm |\bar{x}\rangle)$. Without loss of generality, we will assume $| k^{(0)} \rangle$ is the positive superposition $|+_{x}\rangle$, as is the case for the ground state given our conventions. The overlap between zeroth order eigenstates follows directly from these definitions: $\langle k^{(0)} | k_{MF}^{(0)} \rangle = \frac{1}{\sqrt{2}}$.

To find the overlap between first order corrections, we rewrite these corrections as a sum over degenerate subspaces $D_{m}$ rather than over zeroth order eigenstates. This is possible because of the $\mathbb{Z}_2$ symmetry of each Hamiltonian, allowing the full set of eigenstates to be split into degenerate pairs. 

First, we consider $| k^{(1)} \rangle$. The correction can be expressed abstractly as a weighted sum over states within each degenerate subspace $D_{m}$:
\begin{equation}
\begin{aligned}
    |k^{(1)}\rangle &= \sum_{D_{m} \neq D_{k}} c(m, k) |D_{m} \rangle
    \\ &= \sum_{D_{y} \neq D_{x}} c(y, x) |D_{y} \rangle .
\end{aligned}
\end{equation}
In the second line, the labels are changed to explicitly highlight the relationship to computational basis states $x, y$. Using the previous definition in Eq. (\ref{eq:pt-1st-order-psi}):
\begin{equation}
\begin{aligned}
    c(y, x) |D_{y} \rangle &= \frac{1}{\Delta E_{x, y}} \big( \langle +_{y} | V |+_{x}\rangle |+_{y}\rangle
    + \langle -_{y} | V |+_{x}\rangle |-_{y}\rangle \big)
    \\ &= \frac{1}{\Delta E_{x, y}} \bigg[ \frac{1}{2} \big( V_{yx} + V_{y\bar{x}} + V_{\bar{y}x} + V_{\bar{y}\bar{x}}\big) |+_{y}\rangle
    + \frac{1}{2} \big( V_{yx} + V_{y\bar{x}} - V_{\bar{y}x} - V_{\bar{y}\bar{x}}\big)  |-_{y}\rangle \bigg]
    \\ &= \frac{1}{\Delta E_{x, y}} \big( V_{yx} + V_{y\bar{x}} \big) |+_{y}\rangle , 
\end{aligned}
\end{equation}
where $\Delta E_{x, y}$ is the energy difference between $D_x$ and $D_y$, and $V_{xy} = \langle x | V | y \rangle$ is an element of the matrix $V$ in the computational basis. We have made use of properties of $V$ (namely, $V_{yx} = V_{\bar{y}\bar{x}}$ and $V_{y\bar{x}} = V_{\bar{y}x}$) to simplify.

Rewriting $| k_{MF}^{(1)} \rangle$:
\begin{equation}
\begin{aligned}
    |k_{MF}^{(1)}\rangle &= \sum_{D_{m} \neq D_{k}} c_{MF}(m, k) |D_{m, MF} \rangle
    \\ &= \sum_{D_{y} \neq D_{x}} c_{MF}(y, x) |D_{y, MF} \rangle .
\end{aligned}
\end{equation}
The differing zeroth order states leads to a different expression for the weighted sum:
\begin{equation}
\begin{aligned}
    c_{MF}(y, x) |D_{y, MF} \rangle &= \frac{1}{\Delta E_{x, y}} \big( \langle y | V | x \rangle |y \rangle
    + \langle \bar{y} | V | x \rangle |\bar{y}\rangle \big)
    \\ &= \frac{1}{\Delta E_{x, y}} \big( V_{yx} |y \rangle
    + V_{\bar{y}x} |\bar{y}\rangle \big) .
\end{aligned}
\end{equation}
The inner product between first order corrections is now straightforward: 
\begin{equation}
\begin{aligned}
    \langle k^{(1)} | k_{MF}^{(1)} \rangle &= \sum_{D_{y} \neq D_{x}} 
    c^{*}(y, x) c_{MF}(y, x) \langle D_y | D_{y, MF} \rangle
    \\ &= \sum_{D_{y} \neq D_{x}} \frac{1}{(\Delta E_{x,y})^2} \langle +_{y}| \big( V_{yx} + V_{y\bar{x}} \big) \big( V_{yx} |y \rangle
    + V_{\bar{y}x} |\bar{y}\rangle \big)
    \\ &= \frac{1}{\sqrt{2}} \sum_{D_{y} \neq D_{x}} \frac{1}{(\Delta E_{x,y})^2}  \big( V_{yx} + V_{y\bar{x}} \big)^2
    \\ &= \frac{1}{\sqrt{2}} \langle k^{(1)} | k^{(1)} \rangle .
\end{aligned}
\end{equation}
Finally, the overlap between eigenstates to first order is given by: 
\begin{equation}
\begin{aligned}
    \langle \psi_{k} | \psi_{k, MF} \rangle &= \frac{1}{\sqrt{2}} \mathcal{N}^{*}\mathcal{N}_{MF} \big( \langle k^{(0)} | k^{(0)} \rangle + \langle k^{(1)} | k^{(1)} \rangle \big)
    \\&= \frac{1}{\sqrt{2}} \frac{\mathcal{N}_{MF}}{\mathcal{N}} .
\end{aligned}
\end{equation}

Consider the form of the normalization coefficient $\mathcal{N}$ and $\mathcal{N}_{MF}$:
\begin{equation}
\begin{aligned}
    \mathcal{N} &= \sqrt{\frac{1}{1 +  \langle k^{(1)} | k^{(1)} \rangle}}
    \\ &= \sqrt{\frac{1}{1 + \sum_{D_{y} \neq D_{x}} \big( V_{yx} + V_{y\bar{x}} \big)^2 / (\Delta E_{x,y})^2}}
\end{aligned}
\end{equation}
\begin{equation}
\begin{aligned}
    \mathcal{N}_{MF} &= \sqrt{\frac{1}{1 +  \langle k_{MF}^{(1)} | k_{MF}^{(1)} \rangle}}
    \\ &= \sqrt{\frac{1}{1 + \sum_{D_{y} \neq D_{x}} \big( |V_{yx}|^2 + |V_{y\bar{x}}|^2 \big) / (\Delta E_{x,y})^2}}
\end{aligned}
\end{equation}
In the computational basis, the matrix elements of the perturbing transverse field $V$ are real and all have the same sign (with the sign depending on the sign of $h$). Thus, any product of two matrix elements will be a positive number or zero. Likewise, the squared energy difference $(\Delta E_{x,y})^2$ is positive. The denominator of $\mathcal{N}$ contains two additional factors from the cross-terms of $\big( V_{yx} + V_{y\bar{x}} \big)^2$ that do not appear in $\mathcal{N}_{MF}$; however, these cross-terms vanish as only $V_{yx}$ or $V_{y\bar{x}}$ can be nonzero for the same $y$. It follows that $\mathcal{N}_{MF} = \mathcal{N}$, and:
\begin{equation}
    \langle \psi_{k} | \psi_{k, MF} \rangle = \frac{1}{\sqrt{2}} .
\end{equation}

\begin{figure*}[ht]
    \centering
    \includegraphics[width=5.0in]{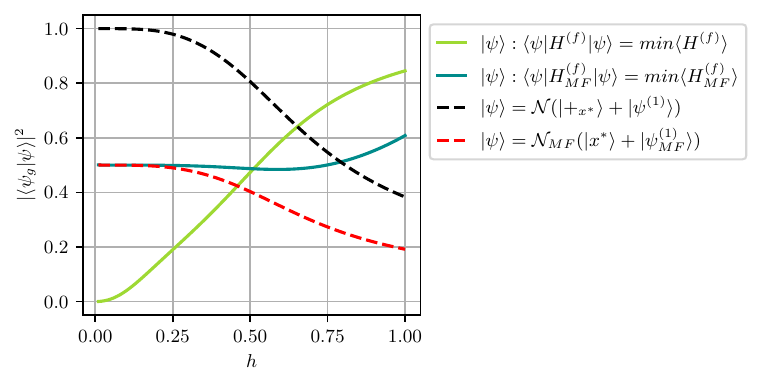}
    \caption{Ising-like model of $N = 6$ with all-to-all $J_{ij}$ sampled from a Gaussian distribution centered at zero with width 1.0. Sweeping across transverse field strength $h$, the ground state $|\psi_g\rangle$ is computed using exact diagonalization of $H$. Fidelity is plotted between $|\psi_{g}\rangle$ and the minimum energy state of $H^{(f)}$ (light green), the minimum energy state of $H_{MF}^{(f)}$ (dark teal), ground state first order perturbation theory for $H$ (black), and ground state first order perturbation theory for $H_{MF}^{(f)}$ (red).}
    \label{fig:PT-example}
\end{figure*}

These results are evaluated numerically for a small system in Fig. \ref{fig:PT-example}. Here, we consider $N = 6$ qubits split into two fragments with $N_f = 3$. The qubits interact all-to-all, with $J_{ij}$ drawn from a Gaussian distribution centered at zero with width 1.0. Three Hamiltonians are considered: the full Hamiltonian $H$, the fragmented Hamiltonian $H^{(f)}$ which neglects all interactions crossing the fragment interface, and finally the mean-field corrected Hamiltonian $H_{MF}^{(f)}$, which replaces interface interactions by mean-field terms. Sweeping across the transverse field strength $h$, the minimum energy state is calculated and compared to the ground state of $H$, $|\psi_g\rangle$. To calculate the minimum energy state of $H_{MF}^{(f)}$ (which is state dependent and thus cannot be computed by solving a linear eigenvalue problem), we perform gradient descent directly on the state vector, minimizing the energy cost. Additionally, the ground states of $H$ and $H_{MF}^{(f)}$ are approximately computed using first order perturbation theory. 

As predicted, the minimum energy state of $H_{MF}^{(f)}$ has fidelity equal to 0.5 with $|\psi_g\rangle$ for small $h$. As $h$ is increased beyond $~0.25$, the system enters a regime where first order perturbation theory is no longer sufficient to describe the state. Finally, we note that for small $h$, the minimum energy state of $H^{(f)}$ has negligible overlap with $|\psi_g\rangle$: when $J_{ij}$ interactions are totally neglected, it is unlikely that the same $|x^*\rangle$ will minimize the energy. 

\section{Batched Pre-training}\label{sec:batch-pretraining}
We employ a batched approach to increase the probability of successful pre-training a PQC. To set up a single pre-training attempt, the full PQC is randomly split into fragments with at most $N_f = 3$ and two auxiliary registers, producing fragmented circuits of size $N_{f+a} = 5$ or smaller, which can comfortably be optimized using classical resources. A number $T = 10$ of such partitioned circuits are generated for the same problem Hamiltonian and optimized in parallel according to Algorithm 1. After their initial optimization, the loss associated with each of the $T$ sets of pre-trained parameters is estimated for the full circuit. This requires $T$ additional loss measurements, a modest overhead when the value $T$ is kept small relative to the required number of iterations. The set of pre-trained parameters producing the smallest loss is selected as the initial starting point for the full circuit optimization.

It is worth mentioning that for large problem Hamiltonians treated using quantum resources, the batched optimization of fragmented circuits can be done classically in parallel prior to using the quantum hardware. Only after pre-training would it be necessary to use the quantum hardware in order to estimate the loss of the pre-trained parameters. The role of batch size $T$ in optimization success is explored numerically in \ref{sec:batch-size}. 

\subsection{The Dependence of VQE Performance on Batch Size}\label{sec:batch-size}
\begin{figure*}[ht]
    \centering
    \begin{minipage}[b]{3in}
    \begin{subfigure}[b]{3in}
        \centering
        \includegraphics[width=3in]{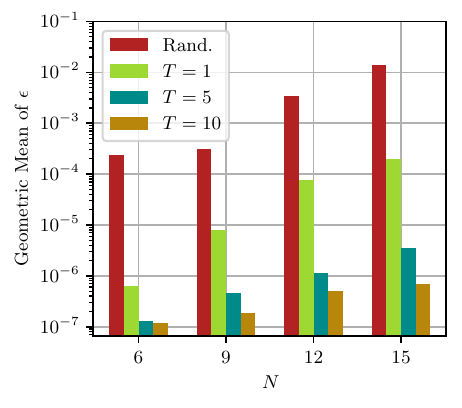}\captionsetup{justification=justified,singlelinecheck=false}
        \caption{}\label{subfig:VQE-avg}
    \end{subfigure}
    \end{minipage}
    \begin{minipage}[b]{3in}
    \begin{subfigure}[b]{3in}
        \centering
        \includegraphics[width=3in]{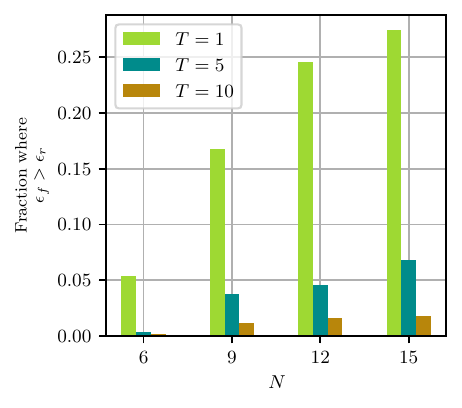}\captionsetup{justification=justified,singlelinecheck=false}
        \caption{}\label{subfig:VQE-comp-fails}
    \end{subfigure}
    \end{minipage}
    \caption{In \ref{subfig:VQE-avg}, the averaged results are presented for three different batch sizes $T$, as well as those of random-initialization. Increasing $T$ reduces the geometric mean of the final percent error, but even a single fragmented pre-trained solution ($T = 1$) provides an orders of magnitude reduction in error on average. In Fig. \ref{subfig:VQE-comp-fails}, a case-by-case comparison of the randomly-initialized solution and the fragmented-initialized solution is considered. The bar height indicates the fraction of the 500 instances considered where fragment-initialization produced a final percent error larger than the randomly-initialized circuit.}
    \label{fig:vqe_batch_size}
\end{figure*}

The role of batch size $T$ is explored in Fig. \ref{fig:vqe_batch_size}. In Fig. \ref{subfig:VQE-avg}, we consider the average performance of fragmented pre-training, contrasting the results to those of vanilla VQE plotted in red. On average, even pre-training a single set of fragmented circuits reduces the final error by orders of magnitude. The performance gain due to fragmented pre-training grows even larger as $T$ is increased, and notably, the amount of error reduction grows with system size. These results indicate that increasing the batch size $T$ will continue to reduce the error on average. In general, larger batch sizes are necessary to produce the same average error as the system size is increased. This can be understood by the fact that the number of possible ways to partition the system can increase with $N$, although exact scaling depends heavily on how the fragmentation is constrained. In Fig. \ref{subfig:VQE-comp-fails}, we consider the fraction of cases where the final percent error of the fragment-initialized optimization is larger than that of vanilla VQE. This corresponds to the number of times that pre-training produced a solution less optimal than that produced by random initialization. Although this fraction is significant when only one set of fragmented circuits is trained (particularly for large system sizes), by increasing $T$ to 5, the fraction is reduced by nearly an order of magnitude.

\end{document}